\documentclass[twocolumn,trackchanges]{aastex631}

\def\msun{\rm{\,M_{\odot}}}

\newcommand{\alerce}{ALeRCE }
\newcommand{\mone}{$m_1\,$}
\newcommand{\mtwo}{$m_2\,$}
\newcommand{\mthree}{$m_3\,$}
\newcommand{\aone}{$a_1\,$}
\newcommand{\atwo}{$a_2\,$}
\newcommand{\btwo}{$b_2\,$}
\newcommand{\logf}{$\log(f)\,$ }

\def\nsample{14 }
\def\nwithNOmone{6 } 
\def\nwithmone{8 } 

\usepackage{amsmath}
\usepackage{cases}

\received{March 5, 2025}

\shorttitle{Peaky Finders}
\shortauthors{Crawford et al.}

\graphicspath{{./}{figures/}}

\begin{document}

\title{Peaky Finders: Characterizing Double-peaked Type IIb Supernovae in Large-Scale Live-Stream Photometric Surveys}

\correspondingauthor{Adrian Crawford}
\email{adrian.crawford@virginia.edu}

\author[0000-0002-7627-4839]{Adrian Crawford}
\affiliation{Department of Astronomy, University of Virginia, 
Charlottesville, VA, 22903, USA}

\author[0000-0001-9227-8349]{Tyler A. Pritchard}
\affiliation{Department of Astronomy, University of Maryland, College Park, MD 20742}
\affiliation{Astrophysical Sciences Division, NASA/Goddard Space Flight Center,
8800 Greenbelt Rd, Greenbelt, MD 20771, USA}
\affiliation{Center for Research and Exploration in Space Science and Technology, NASA/GSFC, Greenbelt, MD 20771}

\author[0000-0001-7132-0333]{Maryam Modjaz}
\affiliation{Department of Astronomy, University of Virginia,
Charlottesville, VA, 22903, USA}

\author[0000-0002-7472-1279]{Craig Pellegrino}
\affiliation{Department of Astronomy, University of Virginia,
Charlottesville, VA, 22903, USA}
\affiliation{Goddard Space Flight Center,
8800 Greenbelt Rd, Greenbelt, MD 20771, USA}

\author[0000-0001-8367-7591]{Sahana Kumar}
\affiliation{Department of Astronomy, University of Virginia,
Charlottesville, VA, 22903, USA}

\author[0009-0004-7268-7283]{Raphael Baer-Way}
\affiliation{Department of Astronomy, University of Virginia,
Charlottesville, VA, 22903, USA}

\begin{abstract}
We present the first photometric population study of double-peaked Type IIb supernovae (SNe IIb). SNe IIb are produced from the core-collapse of massive stars whose outermost Hydrogen layer has been partially stripped prior to explosion. These double-peaked light curves, consisting of a shock-cooling emission peak (SCE) followed by the main nickel-powered peak, contain more crucial information about the progenitor system than the typical single-peaked light curves.  We compiled and analyzed a sample of 14 spectroscopically confirmed SNe IIb---including previously unpublished and re-classified---with publicly available photometric observations, discovered between 2018--2022, from the ZTF and ATLAS surveys. We developed and fit a piecewise linear model, referred to as the ``lightning bolt model,'' to describe the early-time behavior of these objects to measure population statistics. Notably, we find the SCE peak lasts, on average, fewer than five days above half-maximum light with a mean rise time of $2.07\pm1.0$ and $1.1\pm0.8$ mags/day in the g- and r-band respectively. These SCE rise rates are over 10x faster than---and last only a third the duration of---the rise to the nickel-powered peak. These rise times are comparable to those of fast blue optical transient (FBOT) events and we discuss the implications in the text. Finally, we present a proof-of-concept alert filter, using the ANTARES broker, to demonstrate how to translate these population statistics into simple and effective filters to find potential double-peaked SNe IIb in large-scale survey alert streams, like the imminent Vera C. Rubin Observatory Legacy Survey of Space and Time (Rubin LSST).

\end{abstract}

\keywords{Supernovae(1668) --- Core-collapse supernovae (304)}

\section{Introduction} \label{sec:intro}
Some of the most energetic explosions in the universe occur when a massive star ($M > 8 \msun$) ends its life as a core-collapse supernova (CCSN). Observed CCSNe sub-types depend on their progenitor stellar systems, explosion mechanisms, and surrounding environments. Within these CCSNe, some stars will have been stripped of their outermost hydrogen, and sometimes helium, shells and belong to the class called stripped-envelope supernovae (SESNe; \citealt{Filippenko97,GalYam17,Modjaz19}). The mechanisms that cause this stripping/mass loss are an outstanding question. 

Within the core-collapse and stripped-envelope classes lie the Type IIb supernovae (SNe IIb), which are characterized by their spectra which show initially strong hydrogen lines that eventually fade and give way to strong helium lines---suggesting that the progenitor's outer hydrogen layer was partially stripped prior to explosion (e.g., \citet{Woosley94, Modjaz19}). We have directly observed the progenitors of a few of SNe IIb---e.g., 1993J \citep{alderling94, maund09}, 2011dh \citep{Arcavi11, maund11_2011dh}, and 2013df \citep{Van_Dyk_2014}---many of which imply a yellow supergiant progenitor. However, there are still many proposed and viable progenitor channels that can create an empirical Type IIb SN, specifically binary systems and interactions \citep{Smith14, dessart20}. Thus, SNe IIb make for interesting and unique probes of the mass-loss mechanisms that strip stars as well as a probe into the lives and deaths of massive stars. 

The main peak in the optical light curve of most, if not all, SESNe is driven by the radioactive decay of $^{56}$Ni and the shape of this peak is directly related to the ejecta mass (which itself is indirectly related to the progenitor mass) and ejecta velocity \citep{arnett80}, given certain opacity and geometric assumptions. However, some SESNe, including SNe IIb, can show an additional earlier peak. 
This preceding peak is powered by shock-cooling emission (SCE; e.g., \citealt{woosley87, Chevalier92, Woosley94, Richmond96, CandF08, modjaz09, Modjaz19}, for recent reviews see \citet{WandK17, LandN20}) which occurs as the stellar envelope cools after it has been heated by the shock wave which exploded the massive star. The SCE peak is a tracer of the progenitor stellar envelope radius and shock wave geometries \citep{rabinak11, Nakar_2014} if there is no interaction with the circumstellar medium (CSM); it also traces the CSM as shown in \cite{Pellegrino23}. However, recently many instances of CSM interaction have been found---in those cases the shock-cooling emission traces the CSM properties (e.g., \citet{Pellegrino23,Jacobson-Galan22_21gno,Ertini23_21gno}).

The coupling of the SCE peak with the progenitor information from the 
nickel-powered peak makes these double-peaked light curves especially advantageous for stellar forensics. The challenge lies in obtaining photometric observations quickly enough to catch the majority of the SCE peak, particularly in the UV as past work have shown is vital for constraining progenitor models \citep{Pellegrino23}. It is known that the SCE phase of the light curve evolves much quicker than the nickel-powered peak which evolves on a scale of weeks to months (e.g., \cite{Tartaglia_2017, Modjaz19, Ho23}). \cite{Ho23} showed that amongst the fast-evolving transients they found in the ZTF data, the included SNe IIb population showed a range of  time above half-maximum flux, ranging from 3-12 days. 

Prior single-object papers characterizing the shock-cooling emission in individual SNe IIb have used specific, detailed models (each with their own physical assumptions) to describe the object's behavior and possible progenitor channels (e.g. \cite{arcavi_16gkg, Armstrong_2021, Pellegrino23, farah25}). However, vital light curve parameters, such as the observed rise and decline times as well as duration of SCE, have not been quantified in a statistically large sample; this is the main aim of this work. 
With a more data-driven characterization of the evolution of the SCE peak, and of double-peaked SNe IIb light curves in general, we can better study these information-rich objects through statistical samples and more easily identify these objects amongst the transient alert streams. 

While current surveys like Zwicky Transient Facility \citep[ZTF;][]{ztf_bellm}, Asteroid Terrestrial-impact Last Alert System \citep[ATLAS;][]{atlas_tonry}, All-Sky Automated Survey for Supernovae \citep[ASAS-SN;][]{asas-sn}, and Distance Less Than 40 Mpc \citep[DLT40;][]{yang17_dlt40, Tartaglia18_dlt40} have already revolutionized the field, time domain astronomy is about to undergo another transformation with the commencement of the Vera C. Rubin Legacy Survey of Space and Time (Rubin LSST; \citet{Ivezi_2019_LSST}). Rubin LSST is currently predicted to have over a million transient alerts each night, with $\sim 1000$ of those alerts being new supernovae. While brokers will largely handle the brunt of the data influx and management (i.e. by generating alerts) developing quick filters and/or algorithms to find objects of particular interest will a be vital non-broker task. There are already numerous classification architectures and algorithms developed and being developed, each with a specific goal in mind. RAPID focuses on early ID's of transients into one of 12 pre-defined classes \citep{rapid}, Superphot+ provides a redshift-independent classification schema \citep{superphot+}, and \cite{villar21} develops an unsupervised machine learning method to identify anomalous/unexpected transients from the alert stream, just to name a few of the many ongoing classification efforts preparing for the Rubin transient deluge. Many of these classifiers focus on early classification so that we may best allocate our scarce follow-up resources, and be able to trigger and coordinate multi-band and multi-wavelength observations across multiple facilities for our most promising and interesting objects. 

Many of these classifiers are trained on simulated data, to obtain a large enough training and test set, (e.g., \cite[PLAsTiCC;][]{kessler19} and \cite[ELAsTiCC;][]{narayan23}), meaning that any flaws in the theoretical models or simulated dataset are perpetuated through to the classification schema and machine learning features. On the other hand, data-driven classifiers and methods use real, observed data and are not subject to these particular model assumptions; the drawback being the limited sample size. However, in this work we adopt a data-driven approach using \nsample spectroscopically confirmed SNe IIb to limit the numbers of physical assumptions we make while ensuring a representative sample of double-peaked SNe IIb, while keeping the caveats of this approach in mind (see \S\ref{sec:conclusions} for discussion).

In this work we present the largest sample of double-peaked SNe IIb light curves and the first ever early-time population statistics of these objects. While \citet{Khakpash24} included a large sample of SNe IIb light curves, their focus was on analyzing and leveraging publicly accessible SN catalogs, such as the Open SN Catalogue \citep{Guillochon17}; thus, they did not include many SNe IIb with double-peaks besides classical SNe IIb, such as SN 1993J.
Our dataset is described in \S\ref{sec:data}. We create a data-driven, alert-stream-influenced ``lightning bolt'' model to characterize the SCE peak and nickel-powered peak rise, described in \S\ref{sec:model}. From the lightning bolt model we generate population statistics that constrain the SCE rise time, duration, magnitude variation, among many other interesting parameters outlined in \S\ref{sec:fp}. We use these population statistics as features in a proof-of-concept alert stream filter, engineered on the ANTARES broker in \S\ref{sec:filter}, and discuss future improvements and avenues in \S\ref{sec:filter_disc}. Finally, we summarize our methods, findings, and implications of the work in \S\ref{sec:conclusions}.

\section{Description of Dataset}\label{sec:data}
Through a combined literature and Transient Name Server search, we compiled a set of SNe IIb whose data were publicly available survey data, were taken between 2018--2022, consisted of 2 photometric bands, and showed evidence of a double-peaked light curve. This resulted in \nsample objects which are listed in Table \ref{tab:obs_data}.

\begin{deluxetable*}{cccccl}
    \tablecaption{Description of Dataset  \label{tab:obs_data}}
        \tablehead{\colhead{IAU Name} & \colhead{Discovery Name} & \colhead{RA} & \colhead{Dec} & \colhead{z} & \colhead{Paper Reference} } 
        \startdata
         -- & ZTF18aalrxas & 15:49:11.64 & +32:17:16.68 & 0.0582 &\citet{Fremling19}\\ 
         SN~2018dfi & ZTF18abffyqp & 16:50:50.084 & +45:23:52.44 & 0.031302 &\citet{Bruch20, Bruch23}\\
         SN~2019rwd & ZTF19acctwpz &00:10:45.898&+21:08:20.73& 0.017  &\citet{Bruch23}\\ 
         SN~2020ano & ZTF20aahfqpm &13:06:25.176&+53:28:45.53& 0.03113 &\citet{Ho23,Khakpash24}\\
         SN~2020ikq & ATLAS20lfu &13:36:05.016&+28:59:00.11& 0.037 
         &\citet{Ho23,Khakpash24}\\
         SN~2020rsc & ZTF20aburywx &01:19:56.503&+38:11:09.66& 0.0313 &\citet{Ho23,Khakpash24}\\
         SN~2020sbw$^{\rm a}$ & ZTF20abwzqzo & 02:46:03.318 & +03:19:47.66 & 0.023033 & \citet{Bruch23} \\
         SN~2021gno$^{\rm a}$ & ZTF21aaqhhfu &12:12:10.290&+13:14:57.05& 0.006211  &\citet{Jacobson-Galan22_21gno,Ertini23_21gno}\\
         SN~2021heh & ZTF21aaqvsvw &07:59:47.290&+25:21:20.99& 0.026648 & \cite{Soraisam22}\\
         SN~2021pb$^{\rm a}$ & ZTF21aabxjqr & 09:44:46.80 & +51:41:14.6 & 0.033 & \citet{das23} \\
         SN~2021vgn & ZTF21abrgbex &16:21:10.510&+36:03:40.36& 0.032341 & This work \\
         SN~2022hnt & ZTF22aafrjnw &11:36:59.754&+55:09:50.25& 0.0192 & \citet{farah25} \\
         SN~2022jpx & ZTF22aajkpen &10:10:10.000&-11:04:50.05& 0.015  & This work \\
         SN~2022qzr & ATLAS22zpf &00:09:55.001&-05:01:16.09& 0.018705  & This work \\
         \enddata
    \tablenotetext{${\rm a}$}{Typed as Calcium-Rich Transients of Type IIb, see discussion in the text.}
\end{deluxetable*}

Because forced photometry light curves are produced at a specific RA, Dec at every time that there is an observation, these light curves can include additional measurements that may not be included in the alert-stream light curves. For this reason, we choose to utilize forced photometry in our analysis so that we may have the best-sampled, composite-survey light curves possible. 

For this work, we query forced photometry data from the ZTF forced photometry server \citep{ztf_fp} and the ATLAS forced photometry server \citep{atlas_server_smith, atlas_server_shingles} for each object in our sample. We clean the forced photometry light curves by removing bad observations. For ZTF data we apply the quality cuts recommended in \cite{ztf_fp}: {\tt infobitssci}$<33554432$, {\tt scisigpix}$<= 25$, and {\tt sciinpseeing}$\leq4$. Similarly, we apply a reduced chi-squared fit of the PSF to the ATLAS data by selecting observations with {\tt chi/N}$<4$. 
Note, the ZTF forced photometry data products are in flux units while the ATLAS forced photometry data products are in magnitude units. We chose to convert all light curves to AB magnitude space to better match the current ZTF alert-stream. 

We include the discovery, classification, and any relevant details about each object in our sample below. 

\subsection{ZTF18aalrxas}
ZTF18aalrxas was discovered on UT 2018-04-19 07:59:31.2 (JD=2458227.833) by the ZTF survey at J2000.0 coordinates $\alpha$=$15^{h}49^{m}11^{s}.64$, $\delta$=$+32^{\circ}17'16''.8$, at a host-subtracted magnitude of 19.59 in the g-band \citep{19rwd_disc}. It was classified as Type IIb by \citet{Fremling19}.  

\subsection{SN~2019rwd}
SN~2019rwd was discovered on UT 2019-10-05 03:31:40.8 (JD=2458761.647) by the ZTF survey at J2000.0 coordinates $\alpha$=$00^{h}10^{m}45^{s}.888$, $\delta$=$+21^{\circ}08'20''.75$. The discovery was made in the r-band at 18.68 magnitude \citep{19rwd_tns}. It was also detected in the ATLAS and Pan-STARRS surveys. The Transient Name Server (TNS) currently shows it as a Type II \citep{19rwd_tns_type}, however, a recent paper classifies it as Type IIb \citep{Bruch23}. 

\subsection{SN~2020ano}
SN~2020ano was discovered on UT 2020-01-23 11:09:36.0 (JD=2458871.965) by AleRCE/ZTF at J2000.0 coordinates $\alpha$=$13^{h}06^{m}25^{s}.176$, $\delta$=$+53^{\circ}28'45''.53$, in the g-band at 19.097 mag \citep{2020ano_discovery}.  It was classified as Type IIb on TNS by \citet{2020ano_type}.

\subsection{SN~2020ikq}
SN~2020ikq was discovered on UT 2020-04-28 09:47:31.2 (JD=2458967.908) by ATLAS at J2000.0 coordinates $\alpha$=$13^{h}36^{m}04^{s}.998$, $\delta$=$+28^{\circ}58'59''.69$. The discovery was made in the ATLAS c-band at 18.959 mag \citep{20ikq_discovery}. It was also seen in the ZTF and Pan-STARRS surveys. It was classified as a Type IIb on TNS by \citet{20ikq_type}.

\subsection{SN~2020rsc}
SN~2020rsc was discovered on UT 2020-08-19 09:52:06.004 (JD=2459080.9111806) by ALeRCE/ZTF at J2000.0 coordinates $\alpha$=$01^{h}19^{m}56^{s}.503$, $\delta$=$+38^{\circ}11'09''.66$. The first detection was made in the ZTF g-band at 19.5777 mag \citep{20rsc_disc}. It was classified on TNS as a SN IIb by \citet{20rsc_type}. It was also observed by ATLAS.

\subsection{SN~2020sbw}
SN~2020sbw was discovered on UT 2020-08-26 10:43:58.996 (JD=2459087.9472106) by ALeRCE/ZTF at J2000.0 coordinates $\alpha$=$02^{h}46^{m}03^{s}.310$, $\delta$=$+03^{\circ}19'47''.74$ in the r-band at 19.4498 mag \citep{20sbw_disc}. It was also observed by ATLAS and Pan-STARRS. It was classified on TNS as a SN IIb by \citet{20sbw_type} and published as a SN IIb in \cite{Bruch23}. More recently, SN~2020sbw has been categorized as a Type IIb``Calcium-rich" transient in \citet{das23}, belonging to a small but growing group, some of whose members have the spectroscopic class of SNe IIb (see e.g., \citep{De20_CARTclasses}. A number of them (e.g., iPTF 16hgs, SN~2019ehk, SN~2022oqm, see additional SNe in \citealt{das23}) show double-peaked light curves, and while the nature of their progenitors has been debated---whether core-collapse of a low-mass massive star or a white-dwarf explosion---a number of Ca-rich transients of Type IIb, including this object and the other two in our sample (namely SNe~2021gno and 2021pb), have been argued to originate from core-collapse of low-mass massive stars \citep{das23}, at least not fully excluded \citep{Jacobson-Galan20-19ehk,Jacobson-Galan22_21gno,Ertini23_21gno}.

\subsection{SN~2021gno}
SN~2021gno was discovered on UT 2021-03-20 04:48:00.0 (JD=2459293.7) by ZTF at J2000.0 coordinates $\alpha$=$12^{h}12^{m}10^{s}.294$, $\delta$=$+13^{\circ}14'57''.03$ in the r-band at 18.2017 mag \citep{21gno_disc}. It was also detected by ATLAS, Pan-STARRS, and Gaia. TNS shows an initial classification of Type II \citep{21gno_type_ii}, with two subsequent re-classifications of Type IIb and Type Ib \citep{21gno_type_iib, 21gno_type_ib}. However, in more in-depth analyses, SN~2021gno has been categorized as a ``Calcium-rich", or ``Calcium-strong" transient (CaSTs, see detailed analysis in \citealt{Jacobson-Galan22_21gno,Ertini23_21gno}), with \citealt{Ertini23_21gno} not excluding the presence of H. Thus, we include SN~2021gno in our sample of double-peaked SNe IIb, but note that it has low luminosity, as discussed in \S \ref{sec:fbots} and that there is still some controversy whether it was a CCSN or rather a merger of two low-mass White dwarfs \citep{Jacobson-Galan22_21gno}.

\subsection{SN~2021heh}
SN~2021heh was discovered on UT 2021-03-28 05:08:09.6 (JD=2459301.714) by ZTF at J2000.0 coordinates $\alpha$=$07^{h}59^{m}47^{s}.297$, $\delta$=$+25^{\circ}21'20''.97$ in the g-band at 17.51 mag \citep{21heh_disc}. It was also detected by ATLAS, Pan-STARRS, and Gaia. It was typed on TNS as a SN IIb by \citet{21heh_type} and published as such in \cite{Soraisam22}. 

\subsection{SN~2021pb}
SN~2021pb was discovered on UT 2021-01-07 09:30:12.672 (JD=2459221.89598) by ZTF at J2000.0 coordinates $\alpha$=$09^{h}44^{m}46^{s}.816$, $\delta$=$+51^{\circ}41'14''.62$. It was discovered in the g-band at 18.686 mag \citep{21pb_disc}. It was classified as a Type IIb by \cite{21pb_type}. However, SN~2021pb has also been classified as a Calcium-rich SN IIb by \cite{das23}, probably arising from core-collapse of a low-mass massive star.

\subsection{SN~2021vgn}
SN~2021vgn was discovered on UT 2021-08-08 04:29:20.256 (JD=2459434.68704) by ZTF at J2000.0 coordinates $\alpha$=$16^{h}21^{m}10^{s}.513$, $\delta$=$+36^{\circ}03'40''.43$. It was discovered in the r-band at 19.7724 mag \citep{21vgn_disc} and was also detected by ATLAS, Pan-STARRS, and Gaia. TNS currently shows it as a Type IIb  based on our identification \citep{21vgn_type_iib}. 

\subsection{SN~2022hnt}
SN~2022hnt was discovered on UT 2022-04-14 07:03:21.6 (JD=2459683.794) by ZTF at J2000.0 coordinates $\alpha$=$11^36^{m}59^{s}.751$, $\delta$=$+55^{\circ}09'50''.26$ in the g-band at 18.08 mag \citep{22hnt_disc}. It was also detected with ATLAS. It was typed as a IIb on TNS by \cite{22hnt_type, farah25}. 

\subsection{SN~2022jpx}
SN~2022jpx was discovered on UT 2022-05-09 04:55:24.001 (JD=2459708.7051389) 
by ALeRCE/ZTF at J2000.0 coordinates $\alpha$=$10^10^{m}09^{s}.999$, $\delta$=$-11^{\circ}04'50''.00$ in the r-band at 17.3509 mag \citep{22jpx_disc}. It was also detected with ATLAS and Gaia. It was classified as a IIb on TNS by \cite{22jpx_type}. 

\subsection{SN~2022qzr}
SN~2022qzr was discovered on UT 2022-08-09 10:26:23.136 (JD=2459800.93499) by ATLAS at J2000.0 coordinates  $\alpha$=$00^09^{m}55^{s}.006$, $\delta$=$-05^{\circ}01'16''.15$ in the o-band at 18.12 mag \citep{22qzr_disc}. It was also detected with ZTF and Pan-STARRS. It was classified as a Type IIb on TNS by \cite{22qzr_type}.  

We note that the other seven of the nine Ca-rich SNe IIb of \cite{das23} were not included here, as they do not satisfy our photometry quality cuts or did not show two peaks in their light curves.
\section{Modeling the SCE peak and rise to Nickel-powered peak}\label{sec:model}
In order to ascertain whether a new object belongs to a certain population, one must first know what the population looks like on the whole. While the main nickel-powered peak has been well studied and described, the preceding shock-cooling emission peak has proven to be more elusive. Previously published SCE studies of SNe IIb are limited to a single/handful of objects at a time and rely on theoretical models to interpret the SCE peak.
As a complimentary approach, in this work we aim to provide the first population statistics of double-peaked SNe IIb as a class, specifically focused on characterizing the SCE peak's photometric behavior from publicly available alert stream data. Additionally, we provide the largest sample of SNe IIb with SCE to date. 

We aim to use these new population statistics to inform and build a filter that can identify these particular objects from an alert stream. This filter development is discussed more in \S\ref{sec:filter}.

Our lightning bolt model was created with three objectives in mind: one, be able to characterize the early time photometric behavior, specifically the SCE peak; two, generate model-agnostic, data-driven statistics; and three, create easily translatable population statics for use in alert-stream filtering. In regards to the third goal, when a new transient is identified in the alert stream, its light curve consists of very few observations---in the earliest cases, just two detections. This means that any filtering done on an alert stream must involve simple cuts, especially for quickly evolving objects like SCE peaks. The most popular/common filtering criteria are slope (i.e. is this object evolving at a rate that is expected?) and duration (i.e. is this a persisting transient or extremely short-lived transient). To this end, when developing the lightning bolt model, we chose to use three piecewise lines to characterize and model the early-time behavior as their slopes are the most readily translatable into filter cuts. While detailed parabolic or exponential models are more physically motivated and can better inform us of the progenitor characteristics, they require well-sampled light curves for well-constrained fits, and are more computationally expensive.

\begin{figure}
    \centering
    \includegraphics[width=1.0\columnwidth]{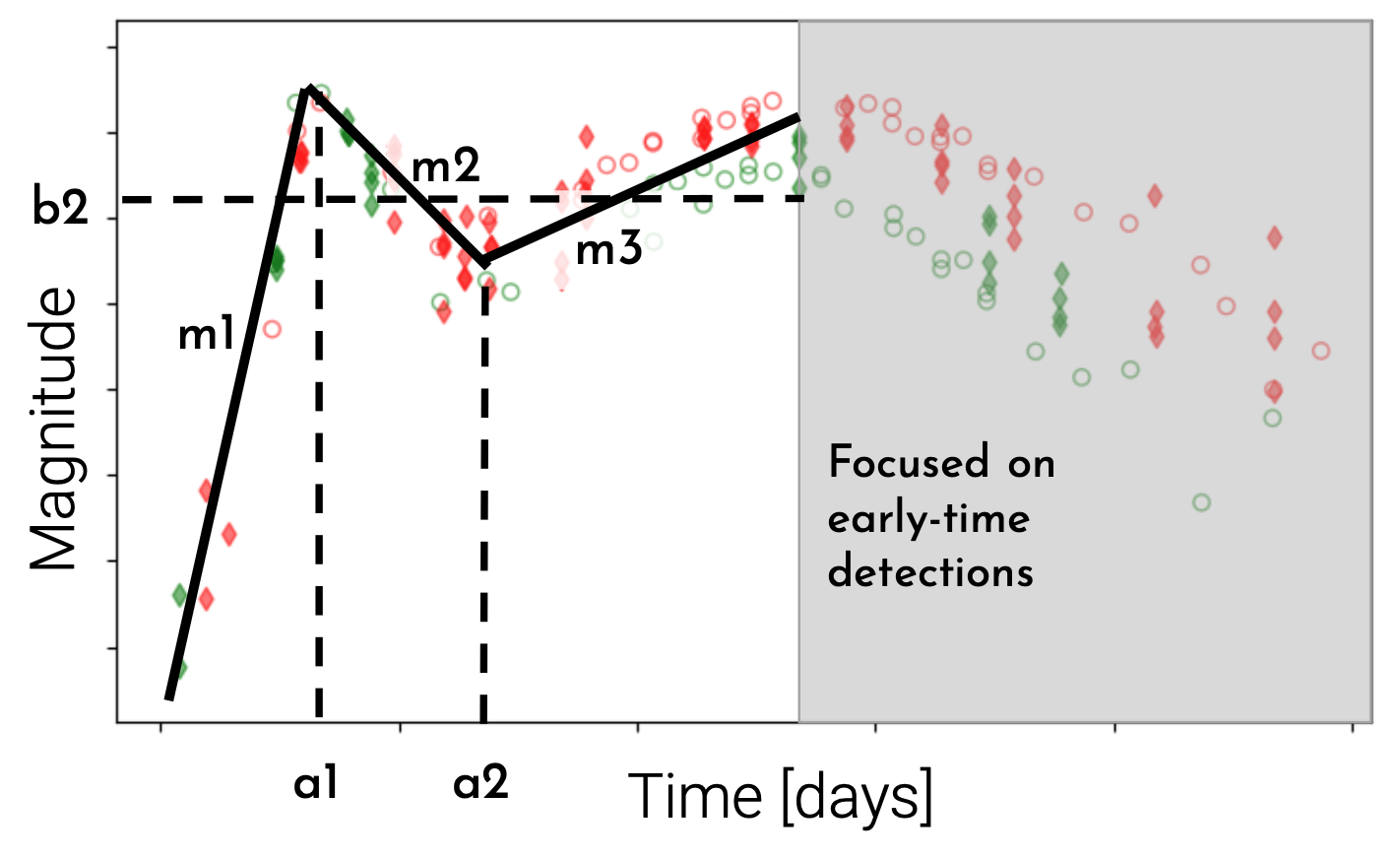}
    \caption{Schematic of the lightning bolt model which is used to fit the early time photometric light curves (stopping before the nickel-powered peak maximum.) The model consists of 7 parameters: \mone the first rise, \mtwo the first decline, \mthree the second rise, \aone the time of the SCE peak, \atwo the time of trough between the two peaks, \btwo the overall magnitude offset, and \logf as estimate of the errors on the errors (not pictured as it is a statistical, not physical, parameter).}
    \label{fig:toy_model}
\end{figure}

With these criteria in mind, we developed the ``lightning bolt'' model, as shown in Figure \ref{fig:toy_model}, which derives its name from the three-line zig-zag shape characteristic of cartoon lightning bolts. Note that the lightning bolt model is only concerned with the earliest photometric behavior and that we do not use or fit observations taken after the nickel-powered peak. 

There are 7 parameters that make up the lightning bolt model:
\begin{itemize}
    \item \textbf{\mone} - slope of the first rise (to SCE peak)
    \item \textbf{\mtwo} - slope of the first decline (from SCE peak)
    \item \textbf{\mthree} - slope of the second rise (to nickel peak)
    \item \textbf{\btwo} - magnitude-axis offset of the model
    \item \textbf{\aone} - time of the SCE peak
    \item \textbf{\atwo} - time of the trough between the SCE and nickel peaks
    \item \textbf{\logf} - estimation of the underestimation of the errors
\end{itemize}

The lightning bolt model takes the parametric form below:

\begin{numcases}{f(x)=}\label{eq:3line}
      m_1x + a_1(m_2-m_1)+b_2 & $x\leq a_1$ \nonumber\\
      m_2x + b_2 & $a_1\leq x\leq a_2$ \nonumber\\
      (m_3x) + a_2(m_2-m_3)+b_2 & $a_2\leq x$ \nonumber
\end{numcases}

However, not all objects have observations comprising the rise to the SCE peak. In these cases, we use a simpler, ``two-line'' version of the lightning bolt that only fits the decline from the SCE peak and the re-rise to the nickel-powered peak, which is described below in Equation \ref{eq:2line}:

\begin{numcases}{f(x)=}\label{eq:2line} 
      \mathrm{{\it NaN}} & $x\leq a_1$ \nonumber\\
      m_2x + b_2 & $a_1\leq x\leq a_2$ \nonumber\\
      (m_3x) + a_2(m_2-m_3)+b_2 & $a_2\leq x$ \nonumber
\end{numcases}

We refer to the original lighting bolt model as the ``full'' model or the ``three-line model'' while the latter is exclusively referred to as the ``two-line model.''

Note that while we apply the same model to each object in our dataset, the initial peak in the double-peaked light curves are not all produced from the same phenomenon or equal progenitor system. As discussed in the introduction, the first peak is sensitive to both shock-cooling physics as well as CSM-interaction (e.g. \cite{Pellegrino23}). Additionally, different physical progenitor channels can produce similar looking light curves, but the similarity in light curve shape does not implicitly imply similar progenitor channels. We do not include any physical modeling or progenitor system analysis. This work's aim is a statistical description of the observed phenomenon of double-peaked light curves in known SNe IIb. We recognize that the inherent variability in progenitor channels and physical environments can be reflected in our sample's measured behavior. Additionally, the quality of our data-driven description is also in part limited by the quality/sampling of our data. 

We apply the lightning bolt model to the forced photometry data of the whole \nsample object sample from ZTF and ATLAS in the following section \S\ref{sec:fp}.

\begin{figure*}
    \centering
    \includegraphics[width=1.85\columnwidth]{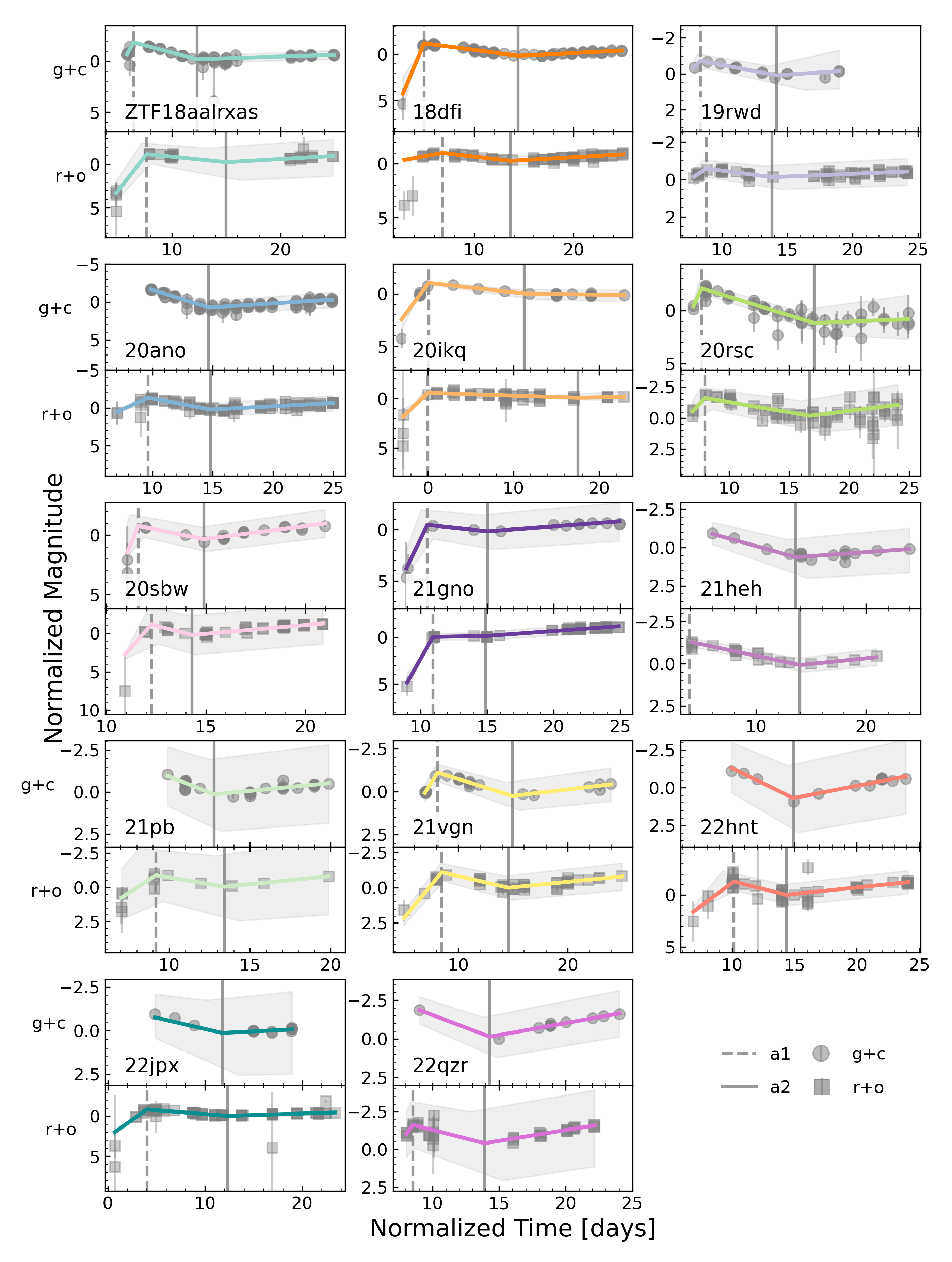}
    \caption{Individual MCMC best-fit models, for each of the 
    \nsample objects' photometry in our sample of double-peaked SNe IIb, are shown in solid colored lines. The solid colored lines are the median best-fit model from the MCMC chains. The gray shaded regions are the $16-84^{\mathrm{th}}$ percentiles. Dashed vertical gray line represents \aone (time of SCE peak) and solid vertical gray line represents \atwo (trough between peaks). Top panels: ZTF g-band and ATLAS c-bands with forced photometry shown in gray circle markers. Bottom panels: ZTF r-band and ATLAS o-bands with forced photometry shown in gray square markers. }
    \label{fig:individ_mcmc_fits}
\end{figure*}

\subsection{Fitting to Forced Photometry Light Curves}\label{sec:fp}
As mentioned in \S\ref{sec:data}, we queried and cleaned the ZTF and ATLAS forced photometry light curves for each of the \nsample objects. While the ZTF forced photometry includes g-, r-, and i-band data, we choose to exclude the data in the i-band for a few reasons: the i-band coverage was too coarse to accurately fit the lightning bolt model, current public ZTF alert streams only utilize g- and r- bands, and ATLAS forced photometry only comes in two bands (c and o). In order to increase the number of observations in each light curve and maintain the two-filter approach, we chose to treat g-band and c-band as one filter and r-band and o-band as another filter. For the remainder of this section, g-band refers to both g and c, and r-band refers to both r and o. The full forced photometry light curves are shown in Figure \ref{fig:full_fp_lcs}.

A brief note on the combination of filters and the $\log(f)$ parameter. As previously described, $\log(f)$ can be thought of as the ``estimation of the underestimation of the errors.''  Meaning, while the observations comprising the light curve all have corresponding errors measurements and while MCMC generates statistically robust error measurements on the model fits, there is still a systematic difference between the two forms of error. One such difference in our case is the treatment of the g- and c-bands as the same filter. While there is significant overlap between the g- and c-band bandpasses, the c-band extends to slightly redder wavelengths than g. The same is true for the r- and o-bands, with o extending further red. We can think of $f$ having units of magnitudes which then allows us to use $\log(f)$ as a measure of the statistical offset between filters. In theory, one could minimize $\log(f)$ with thorough treatment of, and/or conversion between, the ZTF and ATLAS filters; nevertheless, we find our measured statistical uncertainty is typically less than 0.01 mags. (For more details on the mathematical implementation of $\log(f)$ within in the log-likelihood, please refer to the {\tt emcee} documentation: \citealt{FM13}) For the purposes of this project, and thanks to the robustness of the MCMC methodology, we allow $\log(f)$ to encapsulate these subtle systematic error differences.  

To capture the full rise of the SCE we try to include one ``marginal detection'' in each band just before SCE when possible. 
In this work, marginal detections were identified as observations that were significantly dimmer ($\lesssim 4$ mags) than the main SN light curve and had larger errors. 
We chose to only include marginal detections that fell within $\sim 3$ days to the first significant detection in a given band (where significant detections are observations that rise significantly above the baseline/background, generally have smaller errors, and are not isolated from other similar points.) 
In general, our measured slopes on the rise to the SCE peak are lower limits to the possible steepness of the rise.

As mentioned previously, we are only interested in characterizing the early photometric behavior, thus, we exclude any observations that are taken 10 days after $a_2$, i.e. the trough between the peaks. 

In order to describe the population's behavior and shared characteristics we must first normalize the individual objects' light curves.  We chose to shift and align each light curve, in g and r separately, to the trough between the SCE and nickel peak, a.k.a. \atwo in the lightning bolt model. The specific observation corresponding to the trough was visually identified for each object in each band. Each light curve was normalized such that $t(a_2)=15$ days (a fiducial number that allowed nearly all data points to have $t>0$ days) and $\text{mag}(a_2)=0$. 

Since both the two-line and the three-line model include the trough between peaks, we shift all light curves to align on this trough (i.e. \atwo parameter), rather than the SCE peak, i.e. \aone.

After aligning the light curves, we sub-selected the early observations of the light curves, cutting out any observations that fall more than 10 days after the trough. We fit the two-line model to the g-band light curves of SN~2020ano, SN~2021heh, SN~2021pb, SN~2022hnt, SN~2022jpx, and SN~2022qzr. All other g-band light curves and every r-band light curve was fit using the three-line model. 

In order to effectively utilize MCMC to fit the lightning bolt model to light curves, one must properly constrain the parameter spaces and provide a reasonable-enough initial guess. For all parameters, we fine-tuned the bounds on the priors through manual selection of lower and upper bounds followed by inspection of corner plot distributions, ultimately aiming for a gaussian-like distribution centered in the parameter bounds. Finally, we defined our initial guesses by taking the average of the lower and upper bounds of each individual parameter.

We perform the MCMC fitting using the package {\tt emcee} \citep{FM13}, minimizing the log-likelihood, and use the auto-correlation time as a metric for the convergence of the MCMC fits.
We define convergence by a marked flattening in the evolution of the curve of the auto-correlation time. We combine this numeric metric with visual inspection of the corner plots with the best-fit parameter values plotted as cross-hairs, where the intersection of the cross-hairs should fall inside the 1$\sigma$ contour. 
We found that 128 walkers, $2e6$ iterations, and a 500-step burn-in phase achieved convergence for each of the \nsample objects.

In Figure \ref{fig:individ_mcmc_fits}, we show the median best-fit models (solid colored lines) plotted with the 16-84th percentile error band (light gray shaded region) for each model along with the best-fit \aone (dashed vertical gray line) and \atwo (solid vertical gray line) for each object in each band. Note, \nwithNOmone out of \nsample objects do not have any data on the rise to the SCE peak, and thus do not have a best-fit \mone nor \aone. We were able to fit each object very well. 

\begin{figure*}[!tph]
    \raggedright
    \includegraphics[width=.9\columnwidth]{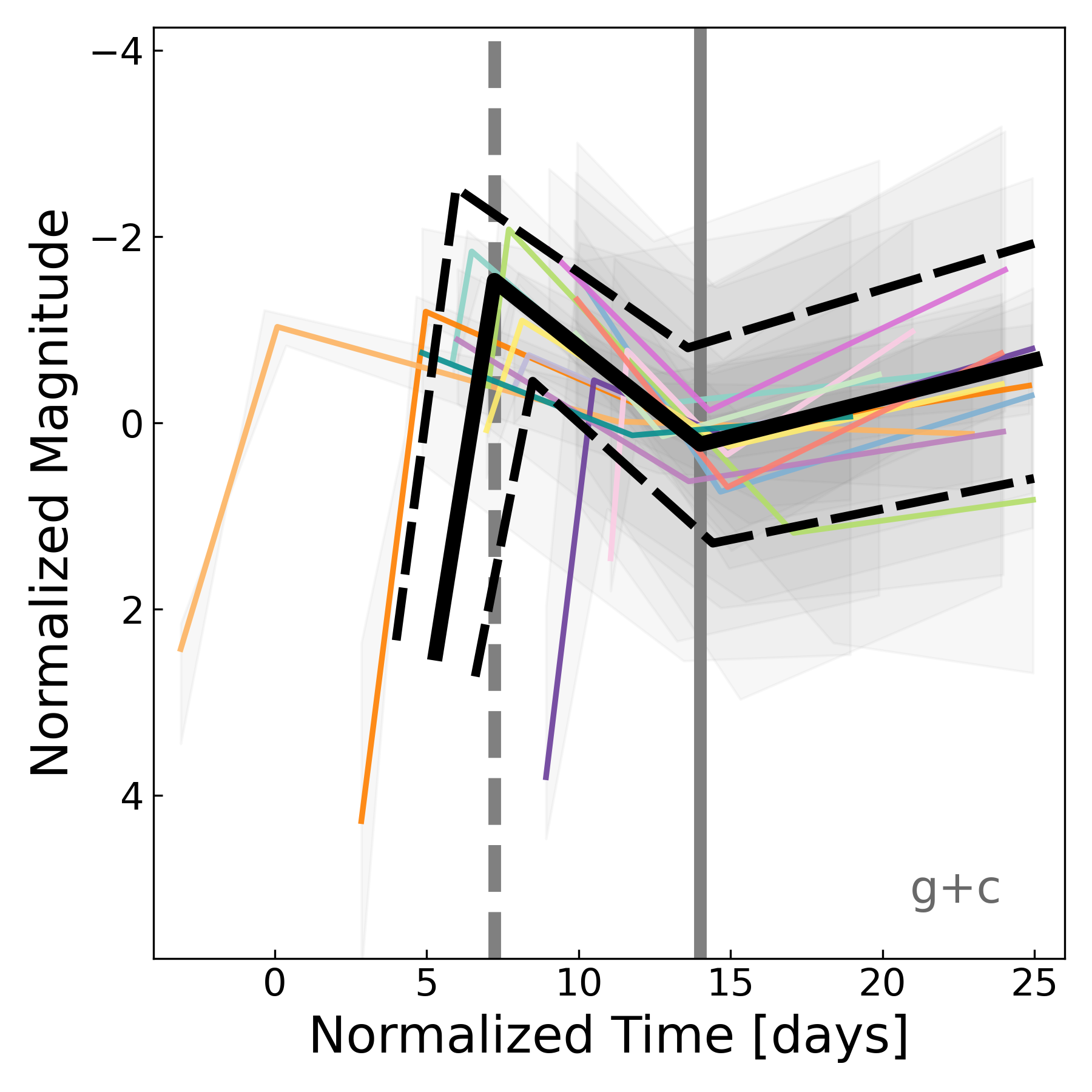}
    \includegraphics[width=1.2\columnwidth]{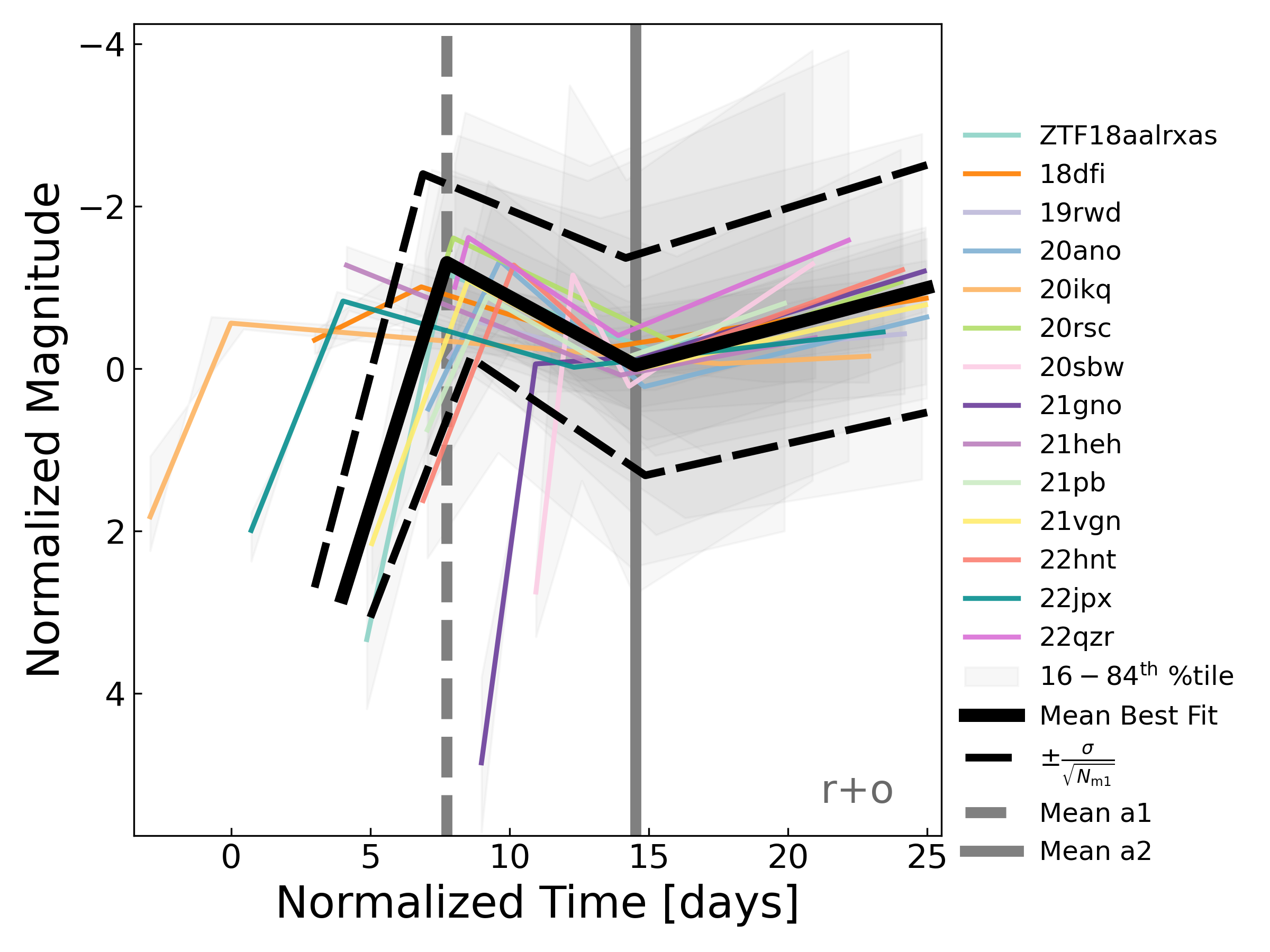}
    \caption{Analysis of the sample's best-fit results using ZTF and ATLAS forced photometry observations. Solid-colored lines are the median fit from the MCMC results with $16-84^{\mathrm{th}}$ percentile error shown in the light gray shaded region. The dashed gray line is \aone and solid gray line is \atwo. The solid black line is the mean of the \nsample best-fit models and we show the standard error of the mean (SEM) in the dashed black lines. Note that while all forced photometry light curves are normalized by eye such that the trough falls at $t=15$ days, the trough from the best-fit MCMC results may fall before, at, or after this fiducial time. The predominant source of scatter between the various light curves is the difference in time between $a_1$ and $a_2$. We see that the slopes of the rise to the SCE peak ($m_1$), while occurring at different relative times, are still quite similar in steepness; the same is true for the slope of the rise to the nickel-powered peak ($m_3$). {\it Left panel}: g+c bands, {\it Right panel}: r+o bands.}
    \label{fig:fp_mean_bestfit}
\end{figure*}

We note for 18dfi that there is a discrepancy in the best-fit $m_1$ slope between the two bands; we acknowledge that this difference in nonphysical and instead due to differences in light curve coverage and number of observations on along $m_1$ which in turn affected the MCMC fits. The shallower r-band $m_1$ slope does not affect the mean of the sample, as presented in Figure \ref{fig:fp_mean_bestfit} and Table \ref{tab:bestfit_stats}. 

In Figure \ref{fig:fp_mean_bestfit} we now show the mean of the \nsample best-fit models in each band along with the standard error of the mean (SEM). We choose to calculate the mean because of the small-number statistics that we are working with. We calculate the mean model fit by taking the mean of each of the parameters and plugging those values into our lightning bolt model, Eq. \ref{eq:3line}. The SEM is shown to quantify the reliability of our mean model rather than the spread in our population.

In the g-band, one object falls outside the SEM region: 20ikq. Looking at its individual plots in Fig. \ref{fig:individ_mcmc_fits}, we see that the time of the SCE peak in both bands occurs around 0 days, whereas most other objects show $5\leq a_1 \leq 10$. This means that the decline from the SCE peak for 20ikq lasted about twice as long as the rest of the sample (see Table \ref{tab:features} for sample decline rate). We see the same slower-than-average evolution in 20ikq in the r-band panel of \ref{fig:fp_mean_bestfit} as well. 


We describe the population statistics from the mean MCMC best-fit parameters in Table \ref{tab:bestfit_stats}. By using the lightning bolt model we were able to obtain rise/fall rates for the three early-time regimes of double-peaked SNe IIb lightcurves. We also show the distribution of the best-fit parameters across the sample in the Appendix in Figure \ref{fig:boxnwhisk}. 
Notably, the rise to the SCE peak (\mone in our model) evolves at a rate of $\approx 2$ mags/day for the g-band and $\approx 1$ mags/day in the r-band. 
In the g-band, this rise to the SCE peak evolves an explosive 25x faster than the rise to the nickel-powered peak. For the r-band, the rise to the SCE peak evolves 12x faster than the nickel-powered rise. Even the decline from the SCE peak (which evolves at around an eighth to a sixth the rate of the SCE rise) still evolves $\approx 3$x faster than the nickel-rise in the g-band and $\approx 2$x faster in the r-band. 
Furthermore, the fall from the SCE peak lasts an average of just seven days. This is a much shorter timescale than the weeks to months that the nickel-powered peak evolves over. We explore this quick SCE evolution timescale further in \S\ref{sec:fbots}.

Looking at Table \ref{tab:features}, we see that the difference in magnitude between the SCE peak and the trough is 1.49 mag in the g-band and 0.96 mag in the r-band. Both of these are easily detectable with current survey limits. Another value of possible interest is the difference in time of the SCE peak maximum between the filters, denoted as $a_1^g-a_1^r$. On average, the peak of the SCE occurs $0.64$ days in the g-band before the r-band. Similarly, when comparing the difference in time of the trough between filters ($a_2^g-a_2^r$), we find that the trough between peaks occurs $0.51$ days in the g-band before the r-band. 

\begin{deluxetable*}{ccccccc}
    \tablecaption{Best-fit population statistics from lightning bolt model \label{tab:bestfit_stats}}
    \tablehead{
        \colhead{Parameter} & \colhead{Units}& \colhead{Min} & \colhead{Mean} & \colhead{Max} & \colhead{StdDev} & \colhead{Filter}}
        \startdata
        $m_1$\tablenotemark{\small $\dagger$} &mags/day     & -4.1	& -2.1 &	-0.7	&1.0     & g  \\
\mtwo &mags/day           & 0.09 &	0.26 &	0.50	&0.12     & g    \\
\mthree &mags/day          & -0.22	&-0.08	&0.01	&0.06     & g   \\
\btwo &mags            & -6.6	&-3.4	&-1.0	&1.7     & g               \\
$a_1$\tablenotemark{\small $\dagger$}& days      & 0.1	&7.2	&11.6	&3.3     & g               \\
\atwo& days             & 11.3	&14.0	&17.1	&1.5     & g               \\
\logf  &...           & -4.8	&-2.3	&0.3	&1.3     & g               \\\hline
\mone  &  mags/day         & -2.95	&-1.10	&-0.17	&0.76     & r         \\
\mtwo   &    mags/day      & -0.02	&0.19	&0.68	&0.16     & r       \\
\mthree  &   mags/day      & -0.23	&-0.09	&-0.02	&0.05     & r     \\
\btwo     &  mags      & -9.5	&-2.8	&0.2	&2.2    & r      \\
\aone     &    days    & 0.0	&7.7	&12.3	&3.1     & r        \\
\atwo     &    days    & 12.3	&14.5	&17.5	&1.3     & r       \\
\logf     &   ...     & -4.6	&-2.4	&-0.7	&1.1     & r     \\ 
        \enddata
    \tablenotetext{\small \dagger}{Values calculated using the \nwithmone (out of \nsample) objects with observations along the rise to the SCE peak. }
\end{deluxetable*}

\begin{deluxetable}{ccccccc}
    \tablecaption{Additional population statistics and features as
derived from best-fit lightning bolt model parameters. \label{tab:features}}
    \tablehead{
        \colhead{Feature\tablenotemark{\small $\rm a$}} & \colhead{Min} & \colhead{Mean} & \colhead{Max} & \colhead{StdDev} & \colhead{N} & \colhead{Units}}
        \startdata
$(a_2-a_1)^g$          & 3.3	&7.0	&11.2	&2.5 & \nwithmone  & days      \\
$(a_2-a_1)^r$          & 2.0	&6.8	&17.6	&3.6  & \nsample   & days    \\
$a_1^g-a_1^r$        & -1.86	&-0.64	&0.10	&0.58  & \nwithmone  & days    \\
$a_2^g-a_2^r$         & -6.3	&-0.5	&0.8	&1.8   &\nsample  & days   \\
mag($a_1$)$^g$ &-2.07 & -1.15 &	-0.45&	0.52 & \nwithmone & mags\\
mag($a_1$)$^r$ & -1.62	& -1.04&	-0.06	&0.41 & \nsample & mags\\
$\Delta\mathrm{mag}(a_2-a_1)^g$   & 0.63	&1.49	&3.26	&0.68	& \nwithmone    & mags   \\
$\Delta\mathrm{mag}(a_2-a_1)^r$     & -0.09	&0.96	&1.55	&0.44	& \nsample  & mags     \\
$\Delta\mathrm{mag}(a_1^g-a_1^r)$    & -0.66	&-0.16	&0.38	&0.29 & \nwithmone & mags      \\
$\Delta\mathrm{mag}(a_2^g-a_2^r)$   & 0.04	&0.37	&1.37	&0.33  & \nsample & mags   \\ 
        \enddata
    \tablenotetext{\small $\rm a$}{Superscripts refer to the band the fitting was performed on. $\Delta$mag$(t_1^{\mathrm{band}}-t_2^{\mathrm{band}})$ refers to the difference in magnitude between the two times.}
\end{deluxetable}

\subsection{Connection to Fast Blue Optical Transients: Peak Luminosity and Timescale of SCE}\label{sec:fbots}

From the results of the previous section, it is clear that the SCE peak evolves on a much shorter timescale than that of the nickel-powered peak. In this section, we further characterize the SCE properties by measuring its absolute magnitude and calculating the time above half-maximum flux ($t_{1/2}$). Further, we compare our results to current FBOT parameter spaces as presented in \cite{Ho23} and find that all \nsample of our objects' SCE peak fall within the typical FBOT timescale of $t_{1/2}<12$ days. 

First, we compute the absolute magnitude of the SCE peak for each object by the following steps. We use a central wavelength of $4830 \AA$  for the g-band and $6260 \AA$ for the r-band. We use the packages {\tt dustmaps} \citep{dustmaps} and {\tt extinction} \citep{extinction} to calculate line-of-sight Milky Way extinction for each SN using the \cite{fm07} formulation. 

To calculate the distance modulus, we first search for published values, which we obtained for ZTF18aalrxas, 21gno, and 21heh. For the remaining object without published distances, we use the redshifts listed in Table \ref{tab:obs_data}. We assume cosmological parameters as outlined in WMAP9 \citep{WMAP9}. 
We use two methods of calculating the absolute magnitude depending on the redshift of the objects. If the object is in the Hubble flow ($z \geq 0.015$ for this work) we use:

\begin{equation}
    M = m - 5\log_{10}\bigg(\frac{D}{10\,\mathrm{pc}}\bigg) + 2.5\log 10(1+z)
\end{equation}

If the object has a redshift $z<0.015$ we use the following instead:

\begin{equation}
    M = m - 2.5\log_{10}\bigg(\frac{D}{10\,\mathrm{pc}}\bigg)^2
\end{equation}

The distances, line-of-sight extinction values, and absolute magnitudes of the SCE peak in both bands are listed in Table \ref{tab:exts}. 

\begin{figure*}[btp]
    \centering
    \includegraphics[width=1.85\columnwidth]{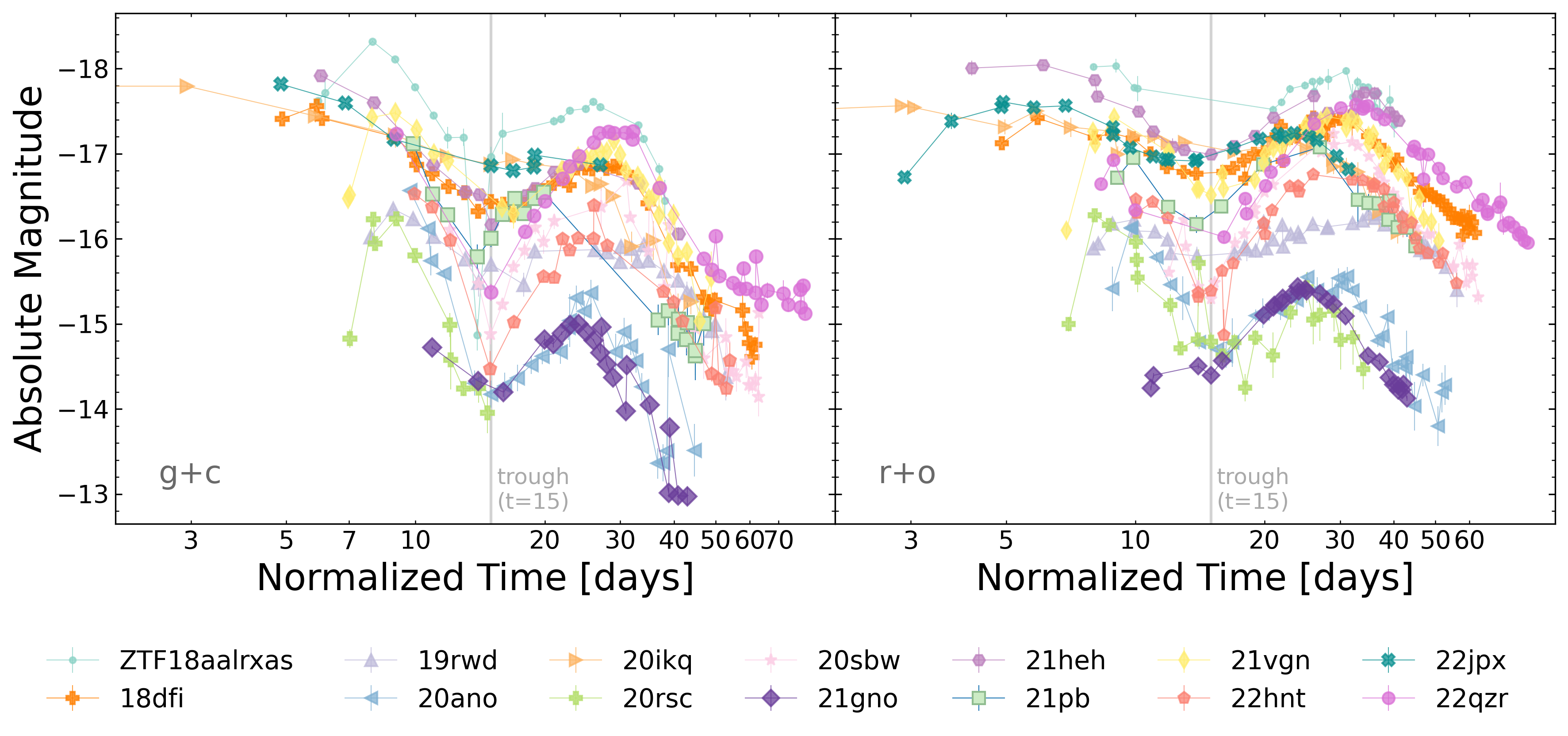}
    \caption{We show the binned light curves of the \nsample objects in absolute magnitude (bin size=2 days; note that light curves shapes may appear different from previous plots due to binning of the ZTF and ATLAS forced photometry together). Each light curve is aligned such that the time of the trough between the SCE peak and the nickel-powered peak falls at 15 days (pale gray vertical line). Time from trough is shown in log-scale. {\it Left panel}: g+c bands, {\it Right panel}: r+o bands.}
    \label{fig:abs_mag_lcs}
\end{figure*}

In Figure \ref{fig:abs_mag_lcs} we show the full light curves in absolute magnitudes (corrected for Milky Way extinction) versus log time to emphasize the early-time photometric behavior. The light curves have been binned for clarity (bin size=2 days) and are aligned such that the trough between the two peaks falls at $t=15$ days. The spread at the trough between peaks spans about 4 magnitudes, from $-14.8$ to $-18.7$ mags in g (for objects fit with the full lightning bolt model) and $-14.4$ to $-18.3$ mags in r for all objects. The mean absolute magnitude at the time of the SCE peak (i.e. $a_1$) was $-17.0\pm 1.2$ and $-16.9\pm 0.97$ mags in the g and r respectively. While the silhouette of the nickel-powered peak remains mostly the same across the sample, the SCE peak takes on a variety of shapes with some SCE peaks evolving slowly and shallowly (e.g., 20ikq, 21heh, 22jpx) while others evolve quickly and sharply (e.g., 20ano, 20rsc, 22qzr). 

We note that the least luminous SN IIb in our sample is SN~2021gno, which has been typed as a calcium-rich transient. While the origin of this class has been debated, it does not appear to skew our population, since the other 2 members of this class in our sample, SN~2021pb and SN~2020sbw are much more luminous (by almost more than 2 mag). 

\begin{deluxetable}{cccccc}
    \tablecaption{Distances, Milky Way line-of-sight extinction values, and absolute magnitudes at SCE peak \label{tab:exts}}
    \tablehead{
        \colhead{SN} & \colhead{D} & \colhead{$A_g$} & \colhead{$A_r$} & \colhead{$M(a_1)^g$} & \colhead{$M(a_1)^r$} \\
        \colhead{} & \colhead{[Mpc]} &\colhead{[Mag]} &\colhead{[Mag]} &\colhead{[Mag]}& \colhead{[Mag]}
        }
        \startdata
        ZTF18aalrxas & 263.0\tablenotemark{$^{\rm a}$} & 0.0244 & 0.0167 & -18.7	& -18.0\\
        SN~2018dfi & 133.2 & 0.0249 & 0.0172& -17.6&	-17.4 \\
        SN~2019rwd & 73.3 & 0.0826 & 0.0575& -16.4	&-16.2 \\
        SN~2020ano & 133.7 & 0.0241 & 0.0167 & {\it NaN} &	-16.2\\
        SN~2020ikq & 162.1 & 0.0134 & 0.00928 & -18.1	& -17.6\\
        SN~2020rsc & 134.4 & 0.0576 & 0.0399 & -16.5	&-16.0\\
        SN~2020sbw & 99.1 & 0.0441 & 0.0306 & -16.2&	-16.6\\
        SN~2021gno &30.5\tablenotemark{$^{\rm a}$} & 0.0417 & 0.0292 & -14.8&	-14.4 \\
        SN~2021heh & 117.7\tablenotemark{$^{\rm a}$} & 0.0416 & 0.0289& {\it NaN}&	-18.3 \\
        SN~2021pb & 142.5 & 0.0130 & 0.00901& {\it NaN}&	-17.0 \\
        SN~2021vgn & 138.9 & 0.0183 & 0.0127 & -17.5	&-17.5\\
        SN~2022hnt & 82.7 & 0.0177 & 0.0123&{\it NaN}&	-16.7 \\
        SN~2022jpx & 64.7 & 0.0552 & 0.0385 & {\it NaN}&	-17.7\\
        SN~2022qzr & 81.8 & 0.0358 & 0.0249 & {\it NaN}&	-17.0\\ 
        \enddata
    \tablenotetext{${\rm a}$}{Taken from the literature, see Table \ref{tab:obs_data} for references.}
\end{deluxetable}

\begin{figure*}[t]
    \centering
    \includegraphics[width=1.75\columnwidth]{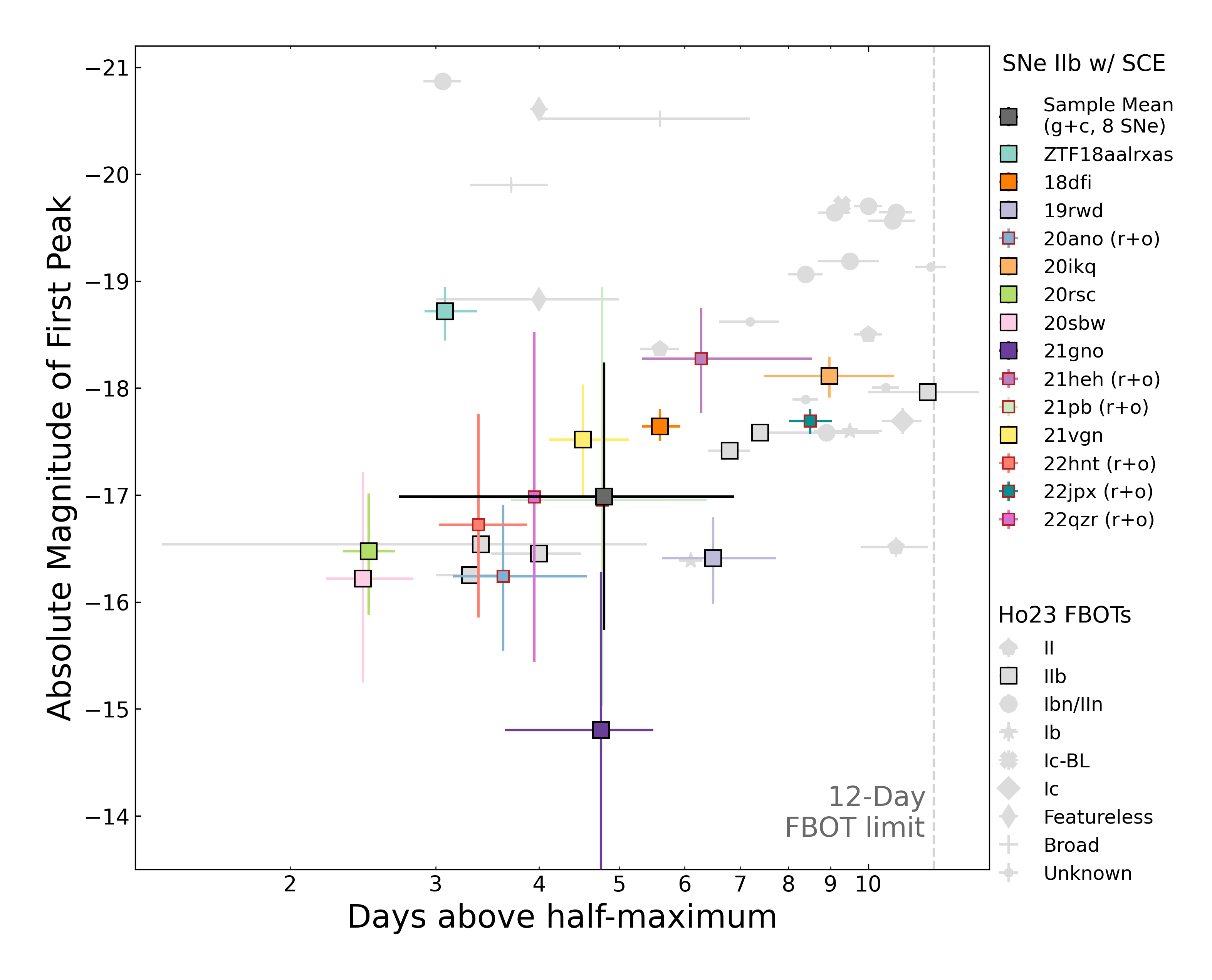}
    \caption{Comparison between the SCE peak absolute magnitude and days above half maximum with literature FBOTs \citep{Ho23}. Pale gray markers represent objects from \citet{Ho23}. Colored square markers represent objects from this work. The black square represents the mean (of the \nwithmone object sample) for the SCE peak (in the $g$+$c$ filters). All objects in this work fell below the 12-day FBOT cutoff, with the sample spending an average of 5 days above half-maximum light. Note, values from \citet{Ho23} are calculated using the g-band. However, not all objects in this work had enough data in the g+c bands to accurately calculate time above half-max. Thus, for the \nwithNOmone objects missing this information, we use the r-band photometry and they are denoted by the boxes with red outlines in the figure. Also note that 3 objects are shared between this work and \citet{Ho23}---20ano, 20ikq, and 20rsc---and the differences between our measurements are described in the text. }
    \label{fig:FBOTs}
\end{figure*}

In order to compare the fast-evolving (but well-known) SCE peak of our objects with the emerging population of fast-evolving transients discovered in new high-cadence surveys whose progenitors and powering sources are debated (see review by e.g.\citealt{Inserra19}) in Figure \ref{fig:FBOTs}, we recreate the Fast Blue Optical Transient (FBOT) parameter space presented in \citet{Drout14, Inserra19, Ho23} of absolute magnitude versus time above half-maximum. The gray points are literature values taken from \citet{Ho23} whose absolute magnitudes were measured in the g-band. We choose to compare to this FBOT sample in particular because 1) it includes six SNe IIb both with and without double-peaked light curves, 2) we share three objects between the samples (SN~2020ano, SN~2020ikq, SN~2020rsc), and 3) it is the first comprehensive survey for FBOTs where the survey actively tried and succeeded in obtaining spectroscopic identifications. We plot the FBOTs of all spectroscopic subtypes in the Ho23 sample, but highlight the SNe IIb in particular with an additional border around their plotting symbol (square). 

We overlay our measurements of the SCE peak for the \nwithmone objects that have data comprising the SCE rise and fall. Our g-band measurements are denoted with the larger colored squares with black edges. For the objects without full SCE data, we instead use the r-band measurements (plotted with as smaller boxes with red outlines). To calculate the error in the time above half-maximum measurement, we use the thinned MCMC posterior chains and calculate a $t_{\mathrm{half}}$ at each iteration. From this distribution of $t_{\mathrm{half}}$ measurements we quote the 50th percentile as the best-fit with the 16th/84th percentiles are the lower/upper bound errors. We repeat for each band, for each object. The errors on the absolute magnitude are taken from the 16th/84th percentile lightning bolt models at their respective SCE peak's (i.e. mag(\aone)). We plot the difference between these $a_1$-magnitudes and the median best-fit model's $a_1$-magnitude as the errors. 

Here we discuss and compare the photometry and the inferred parameters of the three SNe IIb in common between our sample and the Ho23 sample: 20ano, 20ikq, and 20rsc. For 20ano, using the r-band due to lack of rise in the g-band, we measure $t_{1/2}=3.62\,^{+0.95}_{-0.47}$ days which agrees with their $t_{1/2}=3.4\pm2.0$ days. We find a slightly dimmer luminosity at $M=-16.2\pm0.7$ mags compared to their $M=-16.538\pm0.03$ mags which could be attributed to our measurement being done in the r-band and theirs in the g-band; regardless, the magnitudes agree within the errors. For 20ikq, we find $t_{1/2}=8.97\,^{+1.77}_{-1.48}$ days and they find $t_{1/2}=11.8\pm1.8$ days, which agree within the errors; and for SCE peak absolute magnitude we find $M=-18.1\pm0.2$, which agrees with their value of $M=-17.962\pm0.03$. Finally, for 20rsc, we measure $t_{1/2}=2.49\,^{0.19}_{-0.17}$ days compared to their $t_{1/2}=3.3\pm0.3$ days, which do not agree within the errors; our SCE peak absolute magnitude of $M=-16.5\pm0.6$ agrees with their $M=-16.249\pm0.07$. While we find different, notably shorter, values for time above half-maximum for 20ikq and 20rsc, these differences could be due to 1) the use of different forced photometry data and/or 2) different methods of fitting the SCE peak rise and decline. We make use of ZTF {\it and} ATLAS forced photometry data alongside a robust MCMC fitting routine while the Ho23 sample only uses ZTF forced photometry with a Monte-Carlo, linear-interpolation fitting technique. 

The mean values of our \nwithmone SNe IIb with g-band photometry is over-plotted in black with $\pm 1\sigma$ error bars and lies at $\mathrm{Mag}_{g}=-17.02\pm 1.39$ and $t_{1/2}=4.95\pm1.49$ days. 
We can draw a number of important conclusions from Figure~\ref{fig:FBOTs}: the mean as well as the individual SN IIb SCE values in our sample are well within the range of the Ho23 FBOTs and certainly within the 12-day classical FBOT cutoff---both for all subtypes as well as for the SNe IIb, though the latter is not surprising since 3 out of 8 SNe IIb are in common with the Ho23 sample. 
This means that the well-known SCE phenomenon in SNe IIb, having been studied in detail since at least SN~1993J (e.g, \citealt{Schmidt93,Woosley94,Richmond96}, so for more than 32 years, can be found in fast/high-cadence transient searches (see section below~\ref{sec:filter} and in the conclusion for more discussion). However, it also means that this well-known SCE phenomenon in SNe IIb can be mistaken as something new if the main second, Ni-driven peak is not observed/followed due to too short of a time baseline in surveys (see Fig. 1 in Ho23 and section 5.8 in \citet{Khakpash24}). Indeed, \citet{Khakpash24} find that two (SNe IIb~2020rsc and 2020ikq) of the six rapidly evolving SNe IIb from Ho23 appear to be SNe IIb with typical Ni-driven peaks and shock-cooling signature. Usually when people refer to FBOTs or rapidly evolving SNe in the literature, especially theorists (e.g, \citealt{Tsuna25-FBOTtheory}), they refer to truly exotic objects like AT2018cow, which only have one peak and are of high luminosity (see Section 5 in \citealt{Ho23}, not the well-known SCE peaks of SNe IIb. 

The rest of the 3 out of the 6 SNe IIb in the Ho23 sample (namely, SNe IIb 2018gjx, 2019rta, 2020xlt) are not included in our plot because they had only one peak (though the light curve of SN~2020xlt is too limited to fully exclude a 2nd peak; \citealt{Khakpash24})---which is very interesting and does indicate a relatively new kind of phenomenon.  This single-peaked light curve could be due to either, 1) the Ni-peak being very weak/not measurable (i.e. low Ni values produced in the explosion) and the observed first peak is due to the SCE, or 2) the SCE peak is missing, but the first peak is the Ni-driven peak with properties very different from normal SNe IIb Ni-driven peaks. For the latter case, see discussion in section 5.8 in \citet{Khakpash24}, which compares the SNe IIb from Ho23 to the Ni-driven SNe IIb lightcurve template they constructed. 

While we know that the early detection of SESNe with SCE is necessary for obtaining progenitor information from the light curve, early detection is also equally important for planning spectroscopic observations. Even when utilizing ground-based Target of Opportunity (ToO) observations, which are typically obtained within days to weeks, they are often too slow to observe SCE, as seen in Fig~\ref{fig:FBOTs} with the sample mean's incredibly short $\sim 5$ days above half-maximum flux. The rapid time evolution of the SCE peak limits spectroscopy to observing facilities that can execute spectroscopic observations with minimal turnaround time (e.g. \cite{Sravan20}).

\section{Feature Engineering Preliminary ANTARES Filter}\label{sec:filter}
In the upcoming age of the Vera C. Rubin Observatory LSST, there will be an estimated 10 million alerts each night \citep{Ivezi_2019_LSST, lsst_numbers_graham}. There has already been important groundwork laid in the development of transient classifiers, each with their own classification goals as mentioned in the introduction. Most of these classifiers focus on categorizing transients into broad SNe and transient classes, often grouping stripped envelope SNe under a single ``Ibc'' class. Because of the highly diverse nature and behavior of stripped envelope SNe, the Ibc class tends to have a lower purity score than other transient/SN classes, i.e. objects that are not actually Ibc's are misclassified as such thus diluting the overall class behavior (e.g., \cite{rapid, Villar_2020, superphot+}). Part of this low purity can, in some cases, be attributed to the use of simulated lightcurves, such as the Photometric LSST Astronomical Time-Series Classification Challenge \citep[PLAsTiCC;][]{kessler19}, which is known to have unreliable SESNe templates and models, which, when used as a training set for a machine learning algorithm, propagates uncertainty and error into the classification schema itself. Note, this was addressed and improved upon in newer simulated light curves released in the Extended LSST Astronomical Time-Series Classification Challenge \citep[ELAsTiCC;][]{narayan23}. See \cite{Khakpash24} for a detailed comparison of simulated to observed SESNe light curves. 

\begin{figure}
    \centering
    \includegraphics[width=0.9\columnwidth]{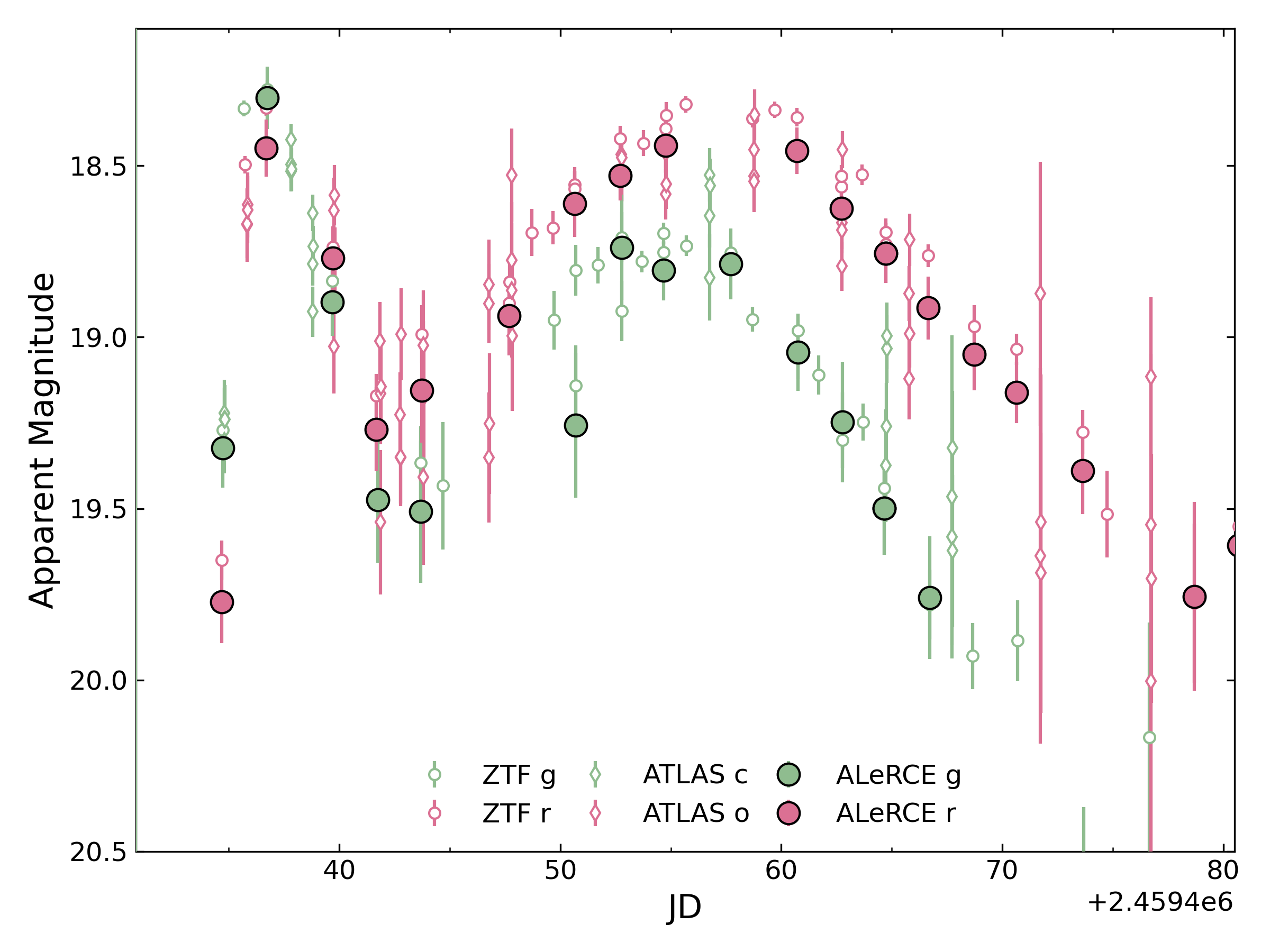}
    \caption{Light curve of SN~2021vgn comparing the coverage between a single alert stream data and multi-survey forced photometry data. The alert stream data is taken from the ALeRCE detections for 21vgn are plotted in the larger circles with black outlines. The forced photometry data is taken from ZTF and ATLAS and are plotted as the smaller empty circles and diamonds respectively. Note that on the rise to the first peak, there are only two detections in each band from the alert data, one close to explosion and one at the SCE peak. In comparison, the combined forced photometry has around 5 observations per band spread from first detection to SCE peak. This light curve sampling disparity will influence how we filter off the alert stream, namely, that broader/more generalized cuts will be prioritized over specific/exact fit measurements. Future time-domain survey designs must prioritize shorter cadences which we approximate here with the multi-survey forced photometry data.
    \label{fig:fp_v_alert}}
\end{figure}

As we saw in the previous section, \S\ref{sec:fp}, the average double-peaked IIb spends an average of 5 days above half-maximum during the first SCE peak; yet the SCE peak contains important and unique probes of SNe IIb progenitors that are not accessible with only the nickel-powered peak, namely, probes of the radius of the progenitor and the mass/radius of the CSM \citep{Soderberg_2012, Morozova_2018, Pellegrino23}. 
Thus, detailed, high-cadence photometry, specifically during the SCE peak, is necessary for the most detailed and insightful analysis of possible progenitors. On top of the detailed photometry, spectroscopy also holds important information that is not contained within the light curves alone.

Our goal, in this classification instance, is to be able to reliably and efficiently identify potential double-peaked IIb SNe from alert streams, such as ZTF and Rubin LSST. Being able to tag potential targets will allow interested astronomers to quickly call for finer-grain photometric follow-up using lesser-subscribed telescopes. 
Additionally, in an age of overwhelming photometric detection and sparing spectroscopic resources, being able to efficiently narrow down the deluge of newly discovered objects into a manageable subset of high-significance objects lets us probe the edges of our physical understanding through targeted and efficient spectroscopic follow-up studies. From a multi-wavelength perspective, fast classification of these objects allows for the triggering of UV telescopes, whose observations during the SCE peak are crucial to tightly constraining the progenitor (and CSM) radius \citep{Pellegrino23}. 

It is no secret that shorter survey cadences lead to better-sampled light curves which in turn are more reliable and descriptive than their sparsely sampled counterparts. Earlier in this work we combine forced photometry from two surveys, ZTF and ATLAS, to create an approximation of a high-cadence light curve for a single source. Now, we contrast those forced photometry light curves with the sparser alert-stream light curves, as seen in Figure \ref{fig:fp_v_alert}. In this figure, we use the \alerce alert stream which runs on the public ZTF survey data and thus has a cadence of 3 days \citep{alerce}. (Note, when developing and testing an alert-stream filter we do so on the ANTARES broker server which also runs on the public ZTF stream; thus, the \alerce and ANTARES alert light curves are effectively identical.)

Filtering for objects off of an alert stream---as is becoming increasingly popular and necessary---presents its own set of challenges. Just looking at the alert light curves of this work's sample, we see a wide spread in the difference in time between an object's first few detections. The shortest length of time between the first two detections of an objects were intranight while the longest was around 2-3 days (i.e. on par with public survey cadence). Thus, when creating filters and features that are specifically designed to be applied to infant light curves with few observations, we must keep this spread in fidelity and cadence in mind. 


The first step in the creation of this classifier was to measure and describe the shared properties (i.e. features) of this class. We did so by utilizing the best-fit parameters from the MCMC fitting as described in the previous sections (\S\ref{sec:fp}). These features are listed in Table \ref{tab:bestfit_stats}. As a reminder, the mean best-fit values quoted in the table are taken from the \nsample median best-fit values of the individual MCMC/model fits. From the initial seven model parameters, we are also able to calculate an additional eight features, as seen in Table \ref{tab:features} which describe the various time, magnitude, and filter offsets between \aone and \atwo. 

We use the mean and standard deviations of the features outlined in Tables \ref{tab:bestfit_stats} and \ref{tab:features} to create a preliminary alert filter on the ANTARES broker \citep{antares}. We do not include $log(f)$ as a parameter as it is a statistical value borne from our MCMC methodology and is not an inherent physical or photometric feature. 
We do not include \btwo as a feature as this parameter described the offset of the whole lightning bolt model in normalized-magnitude units and is not a true physical property of the light curves.
Finally, we do not include \aone or \atwo as individual features as this information is more valuably encoded in the Table \ref{tab:features} features. The creation and testing of this ``zeroth order''/proof-of-concept filter is further described in the following section.

\subsection{``Low-Tech'' Proof-of-Concept ANTARES Filter}\label{sec:filter_v1}
Here the use of ``low-tech'' refers to the lack of machine learning or AI used in the creation and implementation of this version of the alert filter. For our proof-of-concept testing we choose to use historical alert stream light curves (for a subset of this work's 14 object sample), meaning that the light curves of each object have already been observed and contain only public alert stream detections (and non-detections). Thus, the alert stream light curve data can include partial SCE peak evolution and can extend past the peak nickel-powered peak. Note that this filter iteration is the first step in creating a truly ``live-stream'' alert filter that reads in new transient detections as they arrive and does not have access to the full light curve evolution. However, the proof-of-concept filter approach described in this section would easily translate into an archival search tool. 

The high level workflow of the filter is to take an object's alert package (called an ``alert locus'' in the ANTARES vernacular) and pass it through a series of quality checks and feature cuts to ``make sure'' it is a supernova before tallying up the total number of the features from our target sample that are present in the alert object. The output is a 3-tiered tagging system, based on the number of features met/present, ranging from Gold (``Highly probable double-peaked IIb''), Silver (``Probable  double-peaked IIb'') to Bronze (``Potential double-peaked IIb''). If not enough features are met the object is simply skipped/ignored. 

Taking a closer look at each of these overarching steps in the workflow, in order to increase the likelihood that the light curve we are looking at belongs to a supernova we first pass the alert light curve through a series of three checks that ask if:
\begin{itemize}
    \item Alert object is coincident with moving solar system object [if yes, skip object]
    \item There are at least 30 minutes between the first and latest alert detection [if no, skip object]
    \item Alert object is coincident with known stellar object [if yes, skip object]
\end{itemize}

These particular cuts were informed by prior work in detecting fast-evolving targets in transient surveys, namely \cite{Andreoni_2021, Ho_2020}.

If an object's alert light curve passes these initial cuts, we then move onto calculating the 22 possible features describing our sample of double-peaked SNe IIb (as derived from Tables \ref{tab:bestfit_stats} and \ref{tab:features}). First we must identify what shape the light curve has in order to determine the phase and compare to the correct model parameters. We split the alert stream light curve observations into g- and r-band and perform the next steps on each band independently. 
We assign each detection observation with a ``phase'' tag from the following six light curve phases: start, end, rise, fall, peak, and trough. The first detection is automatically labeled {\it start} and the last detection is automatically labeled {\it end}. To determine the slope we use {\tt np.gradient} to calculate the gradient, which uses the next detection ($n+1$) and the previous detection ($n-1$) to determine the slope of a single detection data-point ($n$). 
With each detection being individually labeled with a slope we then assign the {\it rise} and {\it fall} phase tags based on the sign of the slopes. A {\it peak} tag is assigned when the phase tag changes from {\it rise} to {\it fall}, and vice versa for {\it trough}. 
We quickly fit a line to each {\it rise}/{\it fall} phase using {\tt scipy.linregress}, inclusive of {\it peak}/{\it trough} detections. 
With this quick linear regression we can then compare the slope of each phase to our feature-engineered MCMC slope features. We define a minimum/shallowness feature as $\bar{x}-2\sigma$ and a maximum/steepness feature as $\bar{x}+2\sigma$, where $\bar{x}$ is the mean best-fit value from \S\ref{sec:fp}. If the calculated alert slopes meets either of these features (assessed independently), that particular feature is tagged as present in that object. 

After calculating the individual slopes for both bands and assessing if they meet the slope features, we then assess whether the object meets any of the time/magnitude/filter offset features from Table \ref{tab:features}. Because these features have quite a wide $2\sigma$ error band, we only tag a particular feature as present if it meets both the $-2\sigma$ and $+2\sigma$. 

After all the features criteria have been assessed, we add up the number of features present in that object's alert stream light curve. There are 22 total possible features across the two bands. If fewer than 10 features are present, the object is most likely NOT a potential target and is skipped. If 10-11 features are present, we tag the object as ``bronze'' meaning it is potentially a double-peaked IIb SNe. If 12-13 features, we tag the object as ``silver'' meaning this is a probable target object. Finally, if 14 or more features are present we tag the object as ``gold'' meaning there is a strong likelihood that the alert stream light curve belongs to a double-peaked IIb SNe. These cutoffs were determined by optimizing the recovery rate of the alert light curves that had the fullest evolution/coverage. 

We tested the performance of this filter on a set of 10 target objects (i.e. known double-peaked IIb SNe identified in this work: 18dfi, 19rwd, 20sbw, 21gno, 21heh, 21pb, 21vgn, 22hnt, 22jpx, 22qzr) and 33 non-target SNe objects (e.g., SNe II, IIP, Ib, Ic, Ia). We do not include the other 4 objects presented as a part this work in our target sample as their alert light curves did not contain the decline from the SCE peak in at least one band.

Of the 10 target objects, 8 were correctly identified as double-peaked SNe IIb (4 Gold, 1 Silver, and 3 Bronze) while 2 did not meet the required number of features and so were not tagged. Of the 33 non-target objects, 28 were correctly not tagged while 5 were incorrectly tagged as potential double-peaked SNe IIb (1 Silver and 4 Bronze). For the target objects that were (incorrectly) discarded as non-target objects, their alert light curves contained a partial decline from the SCE peak only in a single band. Thus, the total number of features identified, out of the 22 possible sample-derived features, did not meet the minimum of 10 features needed to be tagged. 

In summary, implementing our proof-of-concept filter on historical, non-live-stream light curves taken from the ZTF/ANTARES alert stream resulted in 8 true positives, 28 true negatives, 2 false negatives, and 5 false positives. With these values we can compute the accuracy, completeness, and purity with the following definitions:

\begin{equation}
    \mathrm{accuracy} = \frac{\mathrm{TP+TN}}{\mathrm{TP+TN+FP+FN}}
\end{equation}
\begin{equation}
    \mathrm{completeness} = \frac{\mathrm{TP}}{\mathrm{TP+FN}}
\end{equation}
\begin{equation}
    \mathrm{purity} = \frac{\mathrm{TP}}{\mathrm{TP+FP}}
\end{equation}

The overall accuracy of the filter was 0.837. The completeness was 0.800 and the purity was 0.615. From the initial accuracy and completeness, we see that the filter is adept at identifying both true positives (identifying double-peaked SNe IIb as such) and true negatives (ignoring non-double-peaked SNe IIb). 

\subsection{Discussion of Proof-of-concept Alert Stream Filter}\label{sec:filter_disc}

A potentially powerful improvement to make to the alert stream/survey filter would be the implementation of simultaneous fits across photometric bands---making use of all photometric information at once---rather than splitting the light curves into separate bands. This would allow single-band filtering (currently skipped in this alert filter) and increase flagging information in both fine- and coarse-grain light curves. This combination of multiple filters' information will be especially necessary once the Rubin LSST transient search begins as the current survey strategies will implement a rolling cadence that cycles through multiple filters. 

Another potential improvement would be the development of a larger and more diverse training sample which in turn could be used to create a ``smarter'' machine-learning-based filter. The population statistics presented in this sample are generated from 14 objects in two bands. If one were to include more objects and more filter information, there would be potentially additional, or at least robuster, features to translate into a filter. We plan to explore this avenue in future work. 

Looking back at the completeness and purity of the current iteration of the filter, and how one might improve those metrics, it is important to keep in mind that SNe IIb are not the only objects to evolve quickly or even to produce multi-peaked light curves. Looking again at Figure \ref{fig:FBOTs}, SNe IIb do carve out a somewhat unique parameter space with most objects falling between $-15\leq M \leq -18.5$ and $2\leq t_{1/2} \leq 8$, where most of the object falling within this parameter space belong to the IIb class; however, at the outskirts of this region we see II's, Ib's, Ibn/IIn's and other unknown transients also existing in this parameter space. We know that kilonovae are thought to evolve at similar timescales \citep{Andreoni_2021}. In terms of multi-peaked objects, we know that SNe Ia \citep{kasen10, silverman13, ye24}, SLSNe \citep{gal-yam_slsn_review}, CaRTs \citep{De20_CARTclasses}, and other transients \citep{Soraisam22} all are capable of producing more than a single peak (though the timescales and magnitudes of these additional bumps vary significantly between transient classes.) So, the goal of correctly finding, isolating, and following-up double-peaked SNe IIb specifically is not a simple one. 

When optimizing alert stream filter completeness, it's important to not let the purity suffer, i.e. labeling more non-target objects as targets. Framing this concern in terms of photometric (specifically UV) and spectroscopic follow-up, we cannot afford---both in time and money---to send hundreds of potentially interesting objects to these follow-up facilities; we must be skeptical and restrained. 
While outside the scope of this work, others have begun outlining possible avenues to transition away from human-led decision making and towards automated machine-led decision making (e.g., \cite{Sravan20, Andreoni_2021, Sravan21}) that would need to use as input works, like ours, which are population-derived statistics. 
In an era with millions of transient alerts each night, it would be folly to expect to catch and care about each and every alert, so regardless of completeness, astronomers as a whole must be okay with missing some of our target objects.

\section{Conclusions}\label{sec:conclusions}

\begin{itemize}
    \item We present \nsample spectroscopically confirmed SNe IIb with double-peaked light curves---the largest sample to date---using publicly available survey data from ZTF and ATLAS forced photometry (observed between 2018--2022). Each object's light curve shows an initial peak powered by shock-cooling emission (SCE) which occurs when the shock passes through and heats up the stellar envelope, and a secondary peak powered by the radioactive decay of nickel-56 (i.e., the classical stripped-envelope SNe light curve powering mechanism). 
    
    \item We develop a ``lighting bolt'' model to measure and describe the early-time photometric behavior, specifically focusing on quantifying the rise and fall rates of the SCE peak, which we robustly fit using MCMC. 
    
    \item We generate the first ever early-time photometric population statistics from the mean of the \nsample fits. We find that the rise to the SCE peak evolves 25x faster than the rise to the nickel-powered peak, at an average rate of $\approx 2$ mags/day in the g-band. We find that the decline from the SCE peak lasts an average of one week in both bands. Thus, we suggest that the community use our numbers derived from population statistics to describe and compare to SN IIb SCE peaks, instead of relying on SN~1993J as the "typical" SN IIb with SCE peak.

    \item We measure the time above half-maximum flux ($t_{1/2}$) for the SCE peak for \nwithmone objects in the g-band and \nwithNOmone in the r-band as well as absolute magnitudes, corrected for line-of-sight MW extinction, for all objects. We find that in the g-band, the average $t_{1/2}$ was $\approx 5$ days and that average SCE peaks at $\approx -17$ magnitude. We compare the spread of the SCE from SNe IIb to the latest FBOT parameter space \citep{Ho23} and find that all SCE light curves lie below the 12-day classical FBOT cutoff, with the shortest $t_{1/2}\approx 2.5$ days and the longest $t_{1/2}\approx 9$ days. We interpret our findings to mean that this well-known SCE phenomenon in SNe IIb could be mistaken as an exotic FBOT/fast-evolving transient, if the main second, Ni-driven peak is not observed/followed due to too short of a time baseline in surveys. 

    \item Finally, we present a preliminary framework for a future alert stream filter that implements feature engineering from the lightning bolt model using the ANTARES broker. We test the filter performance on archival light curves, comprising only alert stream data, for 10 target objects (i.e. known double-peaked SNe IIb) and 33 non-target objects. Initial testing resulted in an overall accuracy of 83.7\% and a completeness of 80.0\%.  
\end{itemize}

SNe IIb are important probes into the mechanisms that lead massive stars to lose their outermost envelopes. The double-peaked light curves of SNe IIb are uniquely rich in progenitor tracers as the additional SCE peak holds direct tracers of the progenitor stellar envelope as well as tracers of the shock wave geometries and CSM environment. Past works have shown the importance of observing this SCE peak in the UV, where the emission peaks, and the importance of observations in the UV and optical not just on the decline of the SCE peak but along the rise as well \citep{Pellegrino23}. Thus, observing these objects, especially at early-time during the SCE peak, is very useful for improving our stellar models and better understanding supernova physics. Future work for these objects lies in computing the ratios between the SCE and Ni-powered peaks on a statistically large sample as well as a meta-analysis of all published SNe IIb with double peaks across all bands. 
 
The challenge in finding these important objects lies in their extremely quick evolutionary timescales at the earliest phases of their light curves. We show in this work that the rise to the SCE peak is often missed in modern 2-3 day survey cadences. When it is observed, we find that it evolves 12-25x faster than the ordinary nickel-powered peak and stays above half-maximum flux for just 5 days. We expect these short timescales to only be exacerbated in the UV and bluer filters. 

We see that largest source of photometric difference between the double-peaked SNe IIb light curve shapes in our sample lies in the decay from the SCE peak. Some objects decay quite quickly and sharply from this initial peak (e.g. SN~2020sbw) while others have a shallow and extended decline (e.g. SN~2020ikq). These differences in evolution may exist across a continuum in the photometric SNe IIb behavior, or may (with larger sample sizes) split up into distinct photometric behavioral classes. While this is the largest sample of double-peaked SNe IIb analyzed as a population, it is still small enough to be subject to small-scale statistical effects that influence the overall sample behavior. Additionally, the various SNe IIb light curves in our sample could very well be powered by different physics which in turns reflects in the photometric differences in evolution. In future work we plan to expand the population of spectroscopically confirmed SNe IIb with double-peaked light curves by including historical and private survey data.

In the imminent world of the nightly Rubin LSST transient alert deluge, the already precious spectroscopic resources will become even more overburdened. Even coordinating additional optical photometric follow-up with private survey observing facilities will become harder as simply identifying the objects of interests among the alert stream will become a difficult task. Having a robust picture of a target object's behavior will vastly aid in the search for the needle in the transient haystack. Specifically for Rubin LSST, a single band's light curve will be sparse as Rubin's observing strategy involves a rolling cycle filters. This means that data-driven descriptions that do no rely on highly sampled light curves and that can incorporate multi-filter information will be more powerful tools for identifying objects of interest than models relying on high-cadence single-filter light curves. This need for a broad-strokes, data-driven description of a sample's behavior forms the crux of this work as we use our simple lightning-bolt model to measure and describe the early-time behavior of double-peaked SNe IIb, as a population, for the first time.  

\begin{acknowledgements}
    AC acknowledges support from the Virginia Space Grant Consortium (VSGC) in part from the Graduate Research Fellowship.
M.M. and the METAL group at UVA acknowledge support in part from ADAP program grant 80NSSC22K0486, from NSF grant AST-2206657, and from {\it HST} program GO-16656. 
    TP acknowledges that the material is based upon work supported by NASA under award number 80GSFC24M0006.
    RBW would like to acknowledge support in part from the VSGC Graduate Fellowship.
    The ZTF forced-photometry service was funded under the Heising-Simons Foundation grant \#12540303 (PI: Graham). 
\end{acknowledgements}

\facilities{ATLAS, ZTF}

\software{
ANTARES \citep{antares},
corner \citep{corner},
dustmaps \citep{dustmaps},
emcee \citep{FM13},
extinction \citep{extinction}, 
matplotlib \citep{matplotlib}, 
numpy \citep{numpy}, 
pandas \citep{pandas_software, mckinney-proc-scipy-2010},
scipy \citep{scipy},
}

\bibliography{mybib}{}
\bibliographystyle{aasjournal}

\appendix

\begin{figure*}[h]
    \centering
    \includegraphics[width=0.8\columnwidth]{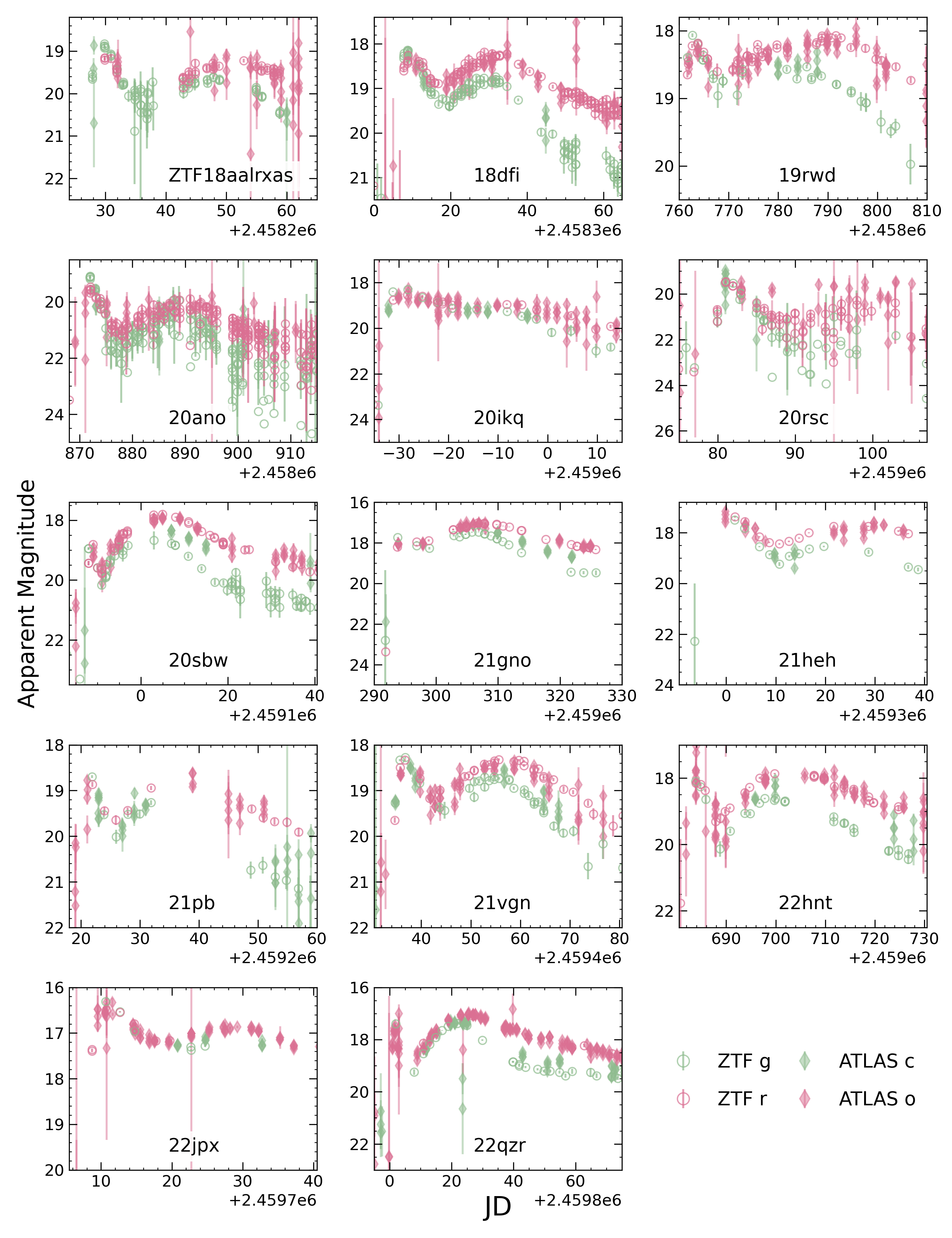}
    \caption{Full forced photometry light curves. ZTF data is shown in the empty circles and ATLAS data is shown in the filled-diamonds. However, note that only the data up to the peak of the nickel-powered peak is used in the MCMC fitting.}
    \label{fig:full_fp_lcs}
\end{figure*}

\begin{figure*}
    \centering
    \includegraphics[width=0.95\columnwidth]{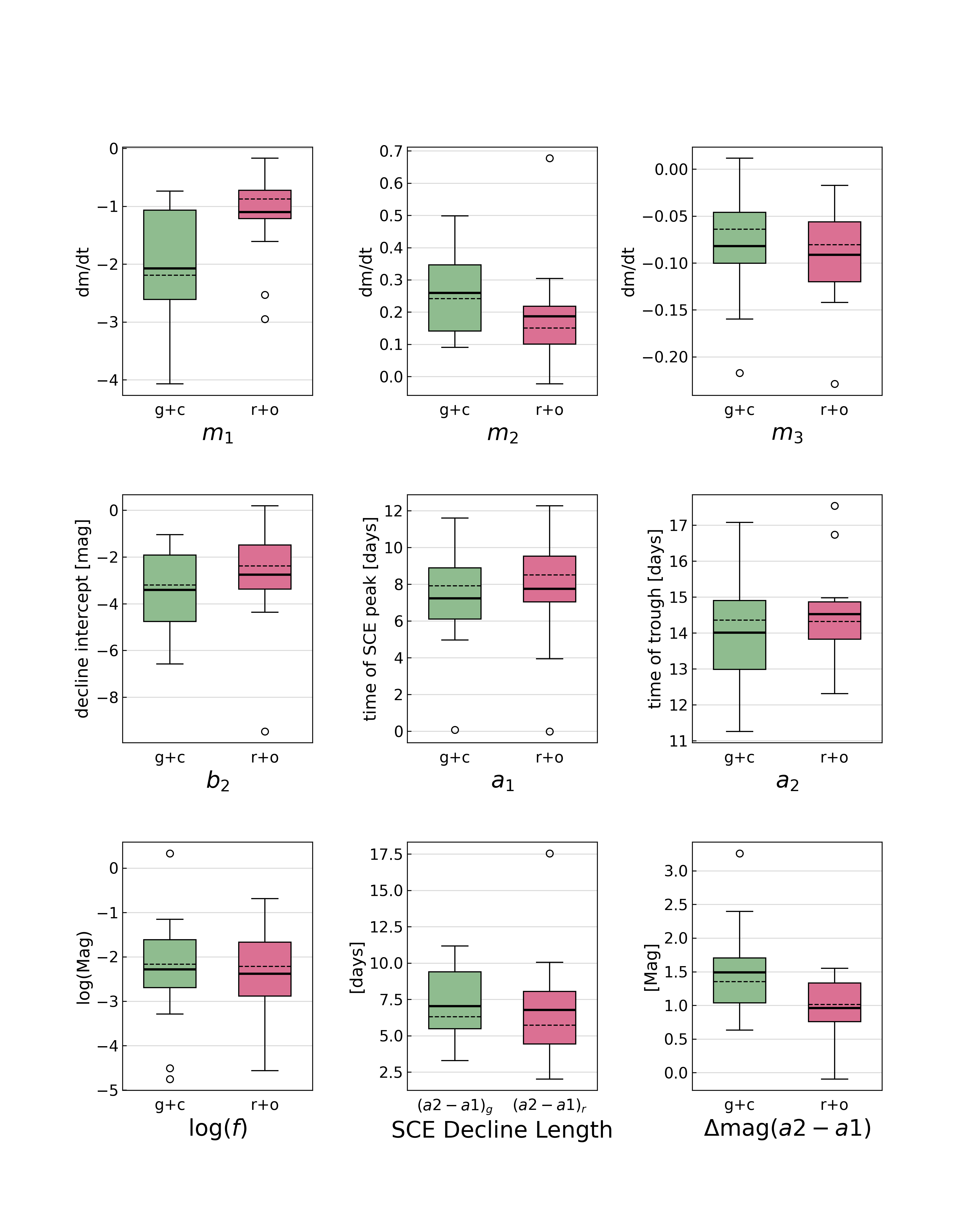}
    \caption{Box-and-whisker plots showing the distribution of best-fit MCMC parameters ($m_1$, $m_2$, $m_3$, $b_2$, $a_1$, $a_2$, $\log(f)$) across the \nsample objects. We also show the distributions of two important population measurements: the duration of the decline from the SCE and the difference in magnitude between the SCE peak and the trough ($\Delta\mathrm{mag}(a_2-a_1)$). The solid, horizontal black line is the population mean and the dashed, horizontal black line is the median. The open circles represent the fliers (i.e. the outliers). }
    \label{fig:boxnwhisk}
\end{figure*}

\end{document}